\documentclass[prd,twocolumn,showpacs,amsmath,amssymb]{revtex4}

\usepackage{graphicx}
\usepackage{dcolumn}
\usepackage{bm}
\usepackage{psfrag}
\usepackage{subfigure}

\newcommand{\be}{\begin{eqnarray}}
\newcommand{\ee}{\end{eqnarray}}
\newcommand{\bn}{\begin{eqnarray*}}
\newcommand{\en}{\end{eqnarray*}}

\newcommand{\nn}{\nonumber \\}
\newcommand{\nl}{\\}

\renewcommand{\th}{\ensuremath{\theta}}
\newcommand{\ph}{\ensuremath{\varphi}}
\newcommand{\al}{\ensuremath{\alpha}}
\newcommand{\bt}{\ensuremath{\beta}}

\newcommand{\lm}{\ensuremath{\lambda}}

\newcommand{\lt}{\ensuremath{\left}}
\newcommand{\rt}{\ensuremath{\right}}

\begin{document}

\pagenumbering{arabic}

\title{The dynamics of a classical spinning particle in Vaidya space-time}%

\author{Dinesh Singh}
\email{singhd@uregina.ca}
\affiliation{%
Department of Physics, University of Regina \\
Regina, Saskatchewan, S4S 0A2, Canada
}%
\date{\today}

\begin{abstract}

Based on the Mathisson-Papapetrou-Dixon (MPD) equations and the Vaidya metric, the motion of a spinning
point particle orbiting a non-rotating star while undergoing radiation-induced
gravitational collapse is studied in detail.
A comprehensive analysis of the orbital dynamics is performed assuming distinct central mass functions
which satisfy the weak energy condition, in order to determine a correspondence between the choice of mass function
and the spinning particle's orbital response, as reflected in the gravitational waves emitted by the particle.
The analysis presented here is likely most beneficial for the observation of rotating solar mass black holes
or neutron stars in orbit around intermediate-sized Schwarzschild black holes undergoing radiation collapse.
The possibility of detecting the effects of realistic mass accretion based on this approach is considered.
While it seems unlikely to observe such effects based on present technology, they may perhaps become observable
with the advent of future detectors.

\end{abstract}


\maketitle

\section{Introduction}

One of the earliest predictions of Einstein's theory of gravitation is the existence
of gravitational waves due to the motion of massive sources.
While strong indirect evidence \cite{Hulse} exists via observations of the Hulse-Taylor binary pulsar
PSR 1913+16--whose orbital period decreases in agreement with general relativity's predictions when
accounting for the emission of gravitational radiation--direct observation of the phenomenon has not yet been achieved.
If gravitational waves exist in the manner predicted by general relativity, then by
a detailed analysis of the shape and amplitude of the generated waveforms, it is theoretically
possible to extract useful information about the physical properties of the collective mass source
that generated the radiation.

The search for gravitational waves is a topic of intense activity, and already a number of
detectors have been built or are under development for the purpose of observing gravitational radiation.
Arguably the most prominent of the gravitational wave detectors in existence or under development
are the ground-based Laser Interferometer Gravitational Wave Observatory (LIGO) \cite{LIGO-ref,LIGO} and the
space-based Laser Interferometric Space Antenna (LISA) \cite{LISA}.
These two facilities complement each other in that LIGO is sensitive to high-frequency gravitational
waves of $10-1000$ Hz generated from the inspiral of equal-mass binary neutron stars or black holes,
while LISA is designed to detect lower frequency waves of $10^{-4}-10^{-1}$ Hz corresponding to the motion of
solar mass neutron stars or black holes in orbit around supermassive black holes.
Other ground-based interferometers that exist worldwide are GEO~600 \cite{GEO}, TAMA~300 \cite{TAMA}, and VIRGO \cite{VIRGO}.
While this form of radiation is notorious for being extremely difficult to detect, gravitational waves offer
a valuable opportunity to provide insight about the sources which generate them.
This includes the earliest moments in the known Universe that are not accessible by other observational means,
since they can propagate over literally cosmological distances with no noticeable dispersion
whatsoever \cite{LISA}.
Gravitational radiation then offers the possibility of gleaning observational information about the
components that drive early Universe cosmology within the first second of the Big Bang,
such as evidence for string interactions and cosmic inflation.
In contrast, information gained from electromagnetic radiation is confined to approximately 300,000 years after the
Big Bang, when temperature fluctuations in the cosmic microwave background radiation can only provide, at best,
some insight about the formation of large-scale structures in the known Universe, with no solid
information about the dynamical processes before the surface of last scattering.

An especially interesting area of general relativity research concerns how ordinary stars of
sufficient mass undergo gravitational collapse to form neutron stars or black holes.
Starting from the first quantitative ``dust'' model introduced by Oppenheimer and Snyder \cite{Oppenheimer},
which assumes a perfect fluid with zero pressure, more sophisticated models \cite{g-collapse}
now exist, each with their own unique features that depend on the star's internal properties.
It appears that gravitational collapse can be categorized into two main types.
The first type concerns stellar core collapse where the central mass remains constant, while
the second type involves the accretion of mass onto the star.
This second form of gravitational collapse has the particularly interesting effect of making the exterior space-time
background {\em dynamic} with the increase of stellar mass, with possible consequences for the motion of any object in its orbit.
Under suitable conditions, such an object may sense a noticeable distortion of the space-time background
that can result in some nontrivial response from an otherwise stable elliptical orbit.
If the orbiting object has sufficient mass, then the gravitational radiation it generates from this dynamical
response may possibly become observable by present or near-future detectors under development.

Another interesting area of general relativity research concerns the dynamics of macroscopic
objects with spin angular momentum propagating in a curved space-time background.
Because most astrophysical objects have acquired at least some spin angular momentum during their formation,
it is very important to know how such objects interact in strong gravitational fields.
This problem was first examined many years ago by Mathisson \cite{Mathisson} who attained a leading order
correction to the geodesic equations in the form of a direct spin coupling to the Riemann curvature tensor.
His work was reformulated by Papapetrou \cite{Papapetrou}, who described the spinning particle in terms of
a matter field confined within a thin world tube in space-time.
Many approaches have been taken to describe the motion of extended objects with spin \cite{Tulczyjew}.
While they identify the leading order correction for the equations of motion, they differ in the contributions
due to higher-order multipole moment expansions of the spinning mass.
Arguably the most conceptually successful model was put forward by Dixon \cite{Dixon},
who proposed a self-consistent set of equations which incorporate to infinite order all multipole moment
expansions of the spinning object.
Regardless, for most practical purposes, truncating the equations of motion to leading order in the spin
interaction is sufficient for a good basic description of the object's physical behaviour in strong
gravitational fields, so long as the spinning particle is small in dimension compared to the space-time's local
radius of curvature.
This is often referred to as the ``pole-dipole approximation,'' first introduced by Mathisson and Papapetrou.
A wide variety of research on classical spinning particles over the past few years have been conducted, which
include both formalism considerations \cite{formalism}, particle kinematics and dynamics in generally
curved space-time \cite{kin-dyn}, and scattering in gravitational wave space-times \cite{scatter}.

Recent applications of these macroscopic spin models have led to some interesting consequences when applied to
black hole space-times.
One example involved studying the motion of the spinning particle's near a Kerr black hole along surfaces of constant azimuthal angle
as defined by the space-time's symmetry axis \cite{Semerak}.
A related example studied the gravitational waveforms generated by a spinning particle while falling into a Kerr black hole
from infinity \cite{Mino} and in circular orbit \cite{Tanaka}.
A particularly interesting example concerned the search for evidence of deterministic chaos within the gravitational
waves generated by a spinning point particle in orbit around a Schwarzschild black hole \cite{Suzuki1,Suzuki2}.
Such a claim would have obvious implications for the ability of existing and future gravitational wave detectors
to extract any meaningful information from the predicted waveforms.
However, a more recent study of the evidence for chaos in a Kerr background suggested that the spins required
to obtain chaotic motion are not likely to be found for astrophysically realistic systems \cite{Hartl1,Hartl2}.

One possible application that has apparently not received any attention so far concerns spinning particle
interaction with a star while undergoing gravitational collapse into a black hole, with the possibility of
observing the phenomenon through gravitational radiation.
This may perhaps be due to the relative unlikelihood of finding a candidate system that captures black hole
formation in progress, compared to fully formed black holes.
Nonetheless, the usefulness of looking into this possibility seems clear, given the great precision required
to generate templates in order to compare with a possible signal.
Because the spinning particle can act like a probe of the space-time background, the gravitational radiation
from such a particle has the ability, in principle, to reflect the dynamical properties of the collapsing star
while in orbit, and that the possibility exists to find signatures of this effect in the gravitational
radiation emitted from it.
\begin{figure}
\psfrag{x}[tc][][2.5][0]{\huge $x$}
\psfrag{y}[tc][][2.5][0]{\huge $y$}
\psfrag{z}[tc][][2.5][0]{\huge $z$}
\psfrag{th}[tc][][2.5][0]{$\hat{\th}$}
\psfrag{ph}[tc][][2.5][0]{$\hat{\ph}$}
\begin{minipage}[t]{0.3 \textwidth}
\centering
\rotatebox{0}{\includegraphics[width = 6.0cm, height = 4.0cm, scale = 1]{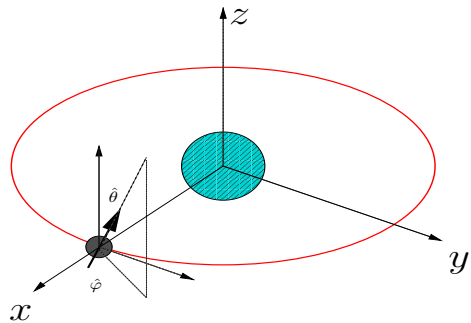}}
\end{minipage}%
\caption{\label{fig:spin-orientation} Initial spin orientation $(\hat{\th},\hat{\ph})$ for the spinning particle.
For the purposes of this paper, the observer is located along the $x$-axis.}
\end{figure}

The purpose of this paper is to examine the dynamical motion of a spinning point particle described
by the Mathisson-Papapetrou-Dixon (MPD) equations in orbit around a non-rotating compact object under gravitational
collapse due to radiation accretion.
The space-time background is then described by the Vaidya metric \cite{Vaidya} in terms of an advanced null co-ordinate
to account for the infall of radiation onto the star's surface.
Although the metric is very simple in form, it nonetheless has valuable properties for cleanly identifying where
the precise effects of gravitational collapse can appear in the $+$ and $\times$ polarization modes,
particularly when the point particle's spin interaction with the background is taken into account.
This paper begins in Section~II with the MPD equations of motion for a spinning point particle,
followed in Section~III by a detailed numerical analysis of the particle's orbit and gravitational radiation
output in the Vaidya background, assuming various mass functions defined by monotonically increasing
elementary functions.
The underlying purpose of Section~IV is to establish the response pattern between the mass function's
growth rate and the orbiting particle's motion, particularly when reflected in the gravitational waveforms.
This is followed by Section~V, which examines the observational possibilities that follow from this
analysis, particularly concerning rotating solar mass black holes and neutron stars in orbit around
intermediate-sized black holes.
The likelihood of detecting the effects of mass accretion based on realistic astrophysical models
is also considered here.
A brief conclusion in Section~VI then completes this paper.
\begin{figure}
\psfrag{t [M]}[tc][][2.5][0]{$t \, [M_0]$}
\psfrag{M(nu) [M]}[bc][][2.5][0]{$M(\nu) \, [M_0]$}
\begin{minipage}[t]{0.3 \textwidth}
\centering
\rotatebox{0}{\includegraphics[width = 6.0cm, height = 4.0cm, scale = 1]{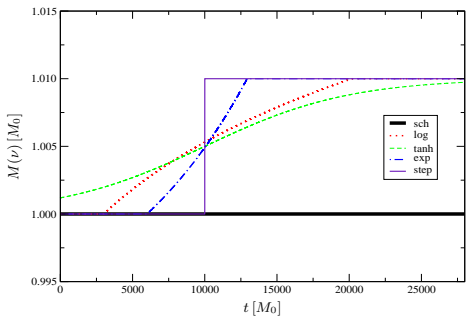}}
\end{minipage}%
\caption{\label{fig:mass-functions} Central mass functions considered for this paper, where $\varepsilon = 0.01$.}
\end{figure}

\begin{figure*}
\psfrag{x[M]}[tr][][1.5][-30]{$\bf x[M_0]$}
\psfrag{y[M]}[bl][][1.5][50]{\hspace{3mm} $\bf y[M_0]$}
\psfrag{z[M]}[][][2][0]{\bf $z[M_0]$}
\begin{minipage}[t]{0.3 \textwidth}
\centering
\subfigure[\hspace{0.2cm} $\lm = 0$]{
\label{fig:sch-orbit:l=0}
\rotatebox{0}{\includegraphics[width = 5.0cm, height = 3.5cm, scale = 1]{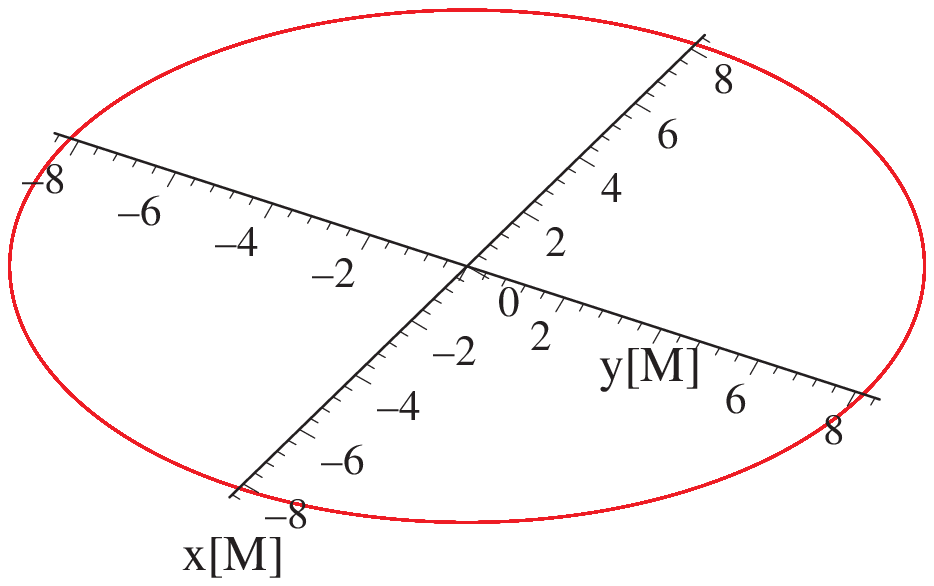}}}
\end{minipage}%
\hspace{0.5cm}
\begin{minipage}[t]{0.3 \textwidth}
\centering
\subfigure[\hspace{0.2cm} $\lm = 1$]{
\label{fig:sch-orbit:l=1}
\rotatebox{0}{\includegraphics[width = 5.0cm, height = 3.5cm, scale = 1]{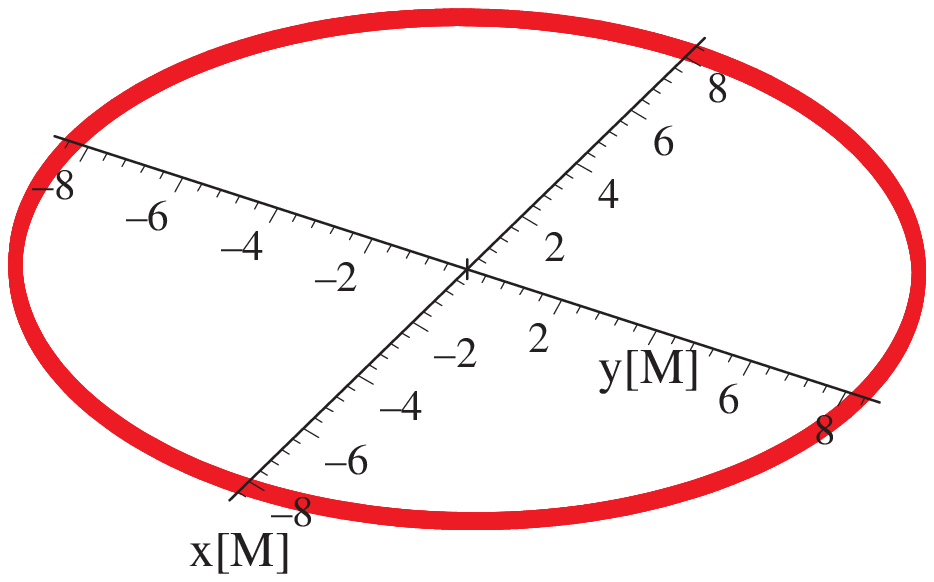}}}
\end{minipage}%
\hspace{0.5cm}
\begin{minipage}[t]{0.3 \textwidth}
\centering
\vspace{-3.5cm}
\caption{\label{fig:sch-orbit} Orbital trajectories for a spinning particle around a Schwarzschild black hole.
Fig.~\ref{fig:sch-orbit:l=0} describes the orbit for a particle without MPD spin-gravity coupling $(\lm = 0)$,
while Fig.~\ref{fig:sch-orbit:l=1} has the coupling included $(\lm = 1)$.
The initial spin orientation is $\hat{\th} = \hat{\ph} = \pi/4$ for both cases.
\vspace{1mm}}
\end{minipage}
\end{figure*}
The space-time metric is described in $+2$ signature, where calculations follow the conventions
adopted in the reference by Misner, Thorne, and Wheeler \cite{MTW}, and geometric units of $G = c = 1$ are assumed throughout.


\section{Equations of Motion for a Spinning Point Particle}

For a spinning test particle, the MPD equations of motion are
\begin{subequations}
\label{MPDeqns}
\be
{DP^\mu \over d\tau} & = & - {1 \over 2} \, R^\mu{}_{\nu \alpha \beta} \, V^\nu \, S^{\alpha \beta} ,
\label{Dp=}
\nl
{DS^{\mu \nu} \over d\tau} & = & P^\mu \, V^\nu - P^\nu \, V^\mu ,
\label{Ds=}
\ee
\end{subequations}
where $V^\mu = dx^\mu/d\tau$ is the particle's centre-of-mass four-velocity defined with respect to some affine parameter $\tau$
that is not necessarily proper time, $P^\mu$ is the particle's four-momentum, and $S^{\mu \nu}$ is the
particle's antisymmetric spin tensor.
The tensor components are defined with respect to a co-ordinate basis.
On their own, (\ref{Dp=})--(\ref{Ds=}) do not completely determine the particle's motion, as they
require two constraints to be satisfied.
The first type is an equation to specify the orthogonality conditions of the spin tensor.
Following Dixon, the choice adopted for the spin condition is $S^{\alpha \beta} \, P_\beta = 0$, which
allows for explicit expression of $V^\mu$ in terms of $P^\mu$ and $S^{\mu \nu}$ \cite{Tod}.
This results in
\be
V^\alpha & = & -{P \cdot V \over m^2} \left[P^\alpha
+ {1 \over 2} \, {S^{\alpha \beta} \, R_{\beta \gamma \mu \nu} \, P^\gamma \, S^{\mu \nu} \over
m^2 + {1 \over 4} \, R_{\mu \nu \rho \sigma} \, S^{\mu \nu} \, S^{\rho \sigma}} \right],
\label{V=}
\ee
where $m = \lt(-P_\mu \, P^\mu\rt)^{1/2}$ is identified as the point particle's mass.
The second type of constraint is a parametrization condition required to determine $P \cdot V$ in (\ref{V=}),
where the choice adopted here is to set $V^0 = {dt/d\tau}~=~1$ throughout the particle's motion.
This allows for identification of the trajectory's time development with the space-time background's co-ordinate time.

The space-time background used to describe gravitational collapse from infalling null shells is
the Vaidya metric \cite{Vaidya}
\be
ds^2 & = & - \left(1 - {2M(\nu) \over r} \right)d\nu^2 + 2 \, d\nu \, dr + r^2 \, d\Omega^2
\label{Vaidya}
\ee
written in Eddington-Finkelstein form, where $d\Omega^2 = d\theta^2 + \sin^2 \theta \, d\varphi^2$
and $\nu  = \nu(t,r)$ is the advanced null co-ordinate.
Given co-ordinates $X^\mu = \lt(\nu, r, \th, \ph \rt)$ corresponding to (\ref{Vaidya}), it follows from
the Einstein field equations that $R_{\mu \nu} = 8 \pi \, T_{\mu \nu}$ for a null fluid source,
where
\be
R_{\mu \nu} & = & \kappa \, L_\mu \, L_\nu,
\label{Ricci}
\nl
\kappa & = & {2 \over r^2} \, {dM(\nu) \over d\nu},
\label{k}
\ee
and $L_\mu = \lt(1, 0, 0, 0 \rt)$ is the ingoing null generator in these co-ordinates satisfying $L_\mu \, L^\mu = 0$.

While it is possible to directly use (\ref{Vaidya}) for the particle motion's time development
with respect to $\nu$, the metric can be expressed in terms of $x^\mu = (t,r,\theta,\varphi)$ co-ordinates
in order to make use of the co-ordinate time $t$.
To do this involves using the tortoise co-ordinate condition
\be
\nu & = & t + r + 2M_0 \, \ln \left({r \over 2M_0} - 1 \right)
\label{nu=}
\ee
to write $d\nu$ as a linear combination of $dt$ and $dr$, where $M_0$ is the static Schwarzschild mass of the central body.
Then by the co-ordinate transformation $X^\mu \rightarrow x^\mu$ and (\ref{nu=}), the Vaidya metric takes the form
\be
ds^2 & = & - \left(1 - {2M(\nu) \over r}\right) \, dt^2
\nn
& &{} + \left(1 - {2 M_0 \over r}\right)^{-1}
\left[2 - {\left(1 - {2M(\nu) \over r}\right) \over \left(1 - {2M_0 \over r}\right)} \right] dr^2
\nn
&  &{} + 2 \left[1 - {\left(1 - {2M(\nu) \over r}\right) \over \left(1 - {2M_0 \over r}\right)} \right] dt \, dr + r^2 \, d\Omega^2
\label{Vaidya-new}
\ee
in terms of lapse and shift vector functions, with the attractive feature that (\ref{Vaidya-new})
reduces to the Schwarzschild metric for $M(\nu) \rightarrow M_0$.
The new ingoing null generator in $(t, r, \th, \ph)$ co-ordinates becomes
\be
l_\mu & = & {\partial X^\al \over \partial x^\mu} \, L_\al
\ = \ \delta^0{}_\mu + \left(1 - {2 M_0 \over r}\right)^{-1} \, \delta^1{}_\mu,
\label{l-mu=}
\ee
and the corresponding Ricci tensor for (\ref{Vaidya-new}) is then
\be
R_{\mu \nu} & = & \kappa \, l_\mu \, l_\nu,
\label{Ricci-new}
\ee
where it can be confirmed that $l_\mu \, l^\mu = 0$.
%


\begin{figure*}
\psfrag{x[M]}[tr][][1.5][-30]{$\bf x[M_0]$}
\psfrag{y[M]}[bl][][1.5][50]{\hspace{3mm} $\bf y[M_0]$}
\psfrag{z[M]}[][][2][0]{\bf $z[M_0]$}
\begin{minipage}[t]{0.3 \textwidth}
\centering
\subfigure[\hspace{0.2cm} log ($\lm = 0$)]{
\label{fig:log:l=0}
\rotatebox{0}{\includegraphics[width = 5.0cm, height = 3.5cm, scale = 1]{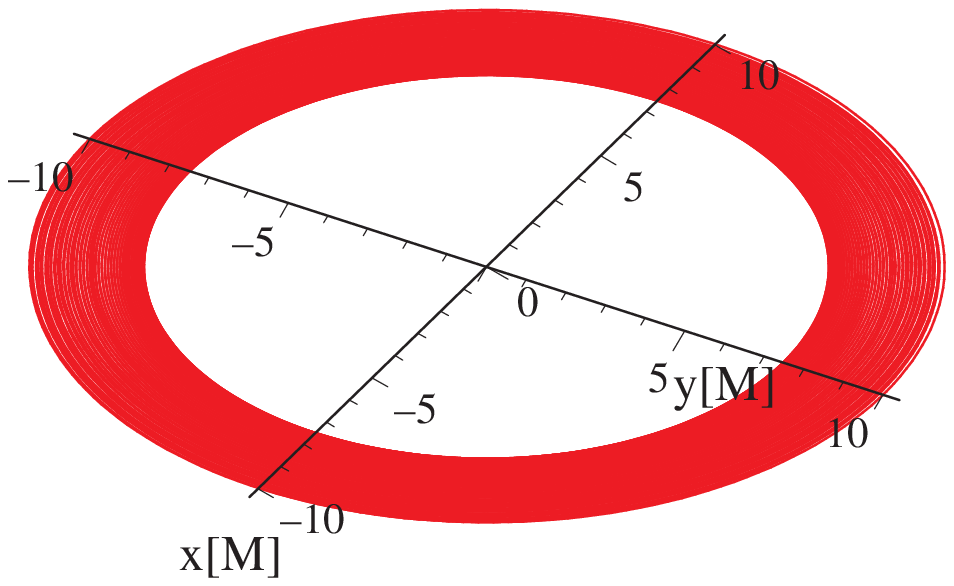}}}
\end{minipage}%
\hspace{0.5cm}
\begin{minipage}[t]{0.3 \textwidth}
\centering
\subfigure[\hspace{0.2cm} tanh ($\lm = 0$)]{
\label{fig:tanh:l=0}
\rotatebox{0}{\includegraphics[width = 5.0cm, height = 3.5cm, scale = 1]{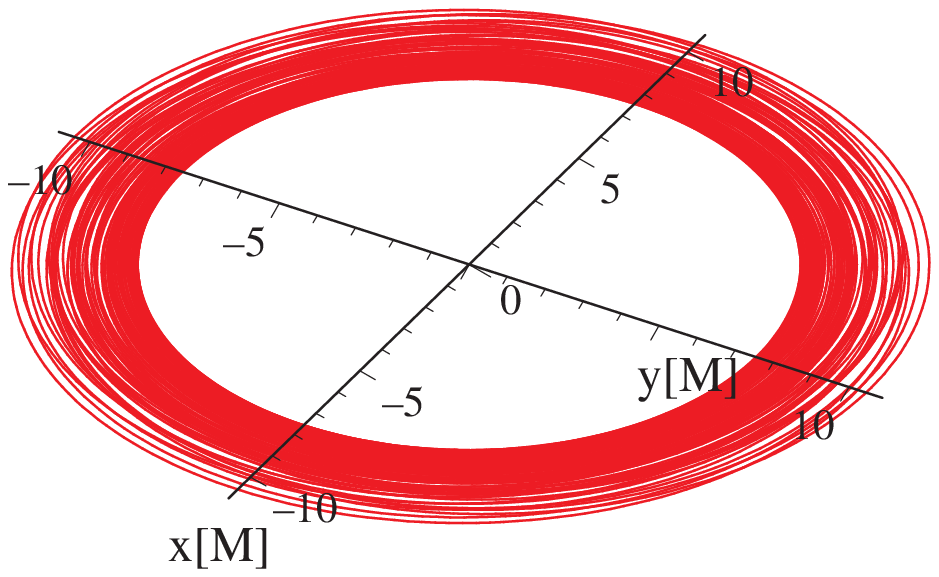}}}
\end{minipage}%
\hspace{0.5cm}
\begin{minipage}[t]{0.3 \textwidth}
\centering
\subfigure[\hspace{0.2cm} exp ($\lm = 0$)]{
\label{fig:exp:l=0}
\rotatebox{0}{\includegraphics[width = 5.0cm, height = 3.5cm, scale = 1]{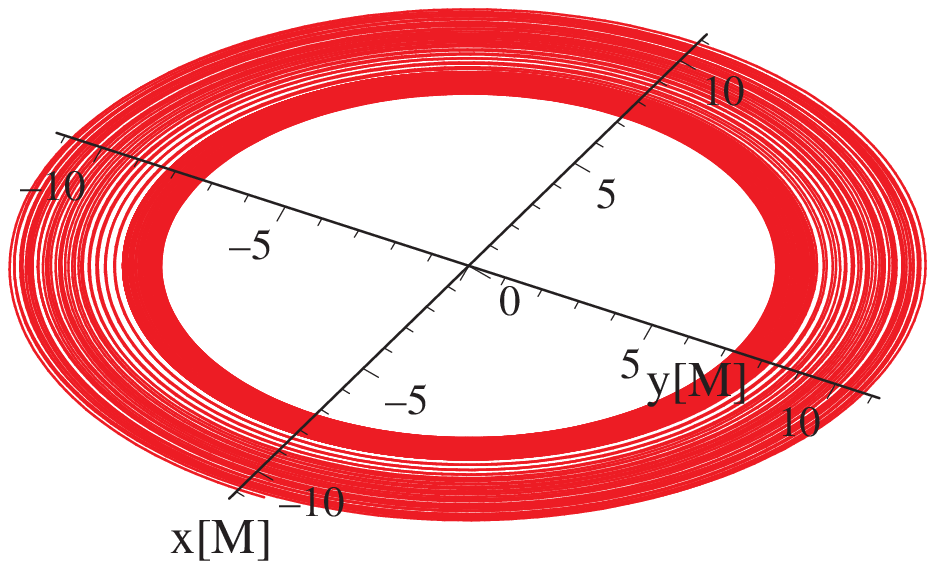}}}
\end{minipage}
\hspace{0.5cm}
\begin{minipage}[t]{0.3 \textwidth}
\centering
\subfigure[\hspace{0.2cm} log ($\lm = 1$)]{
\label{fig:log:l=1}
\rotatebox{0}{\includegraphics[width = 5.0cm, height = 3.5cm, scale = 1]{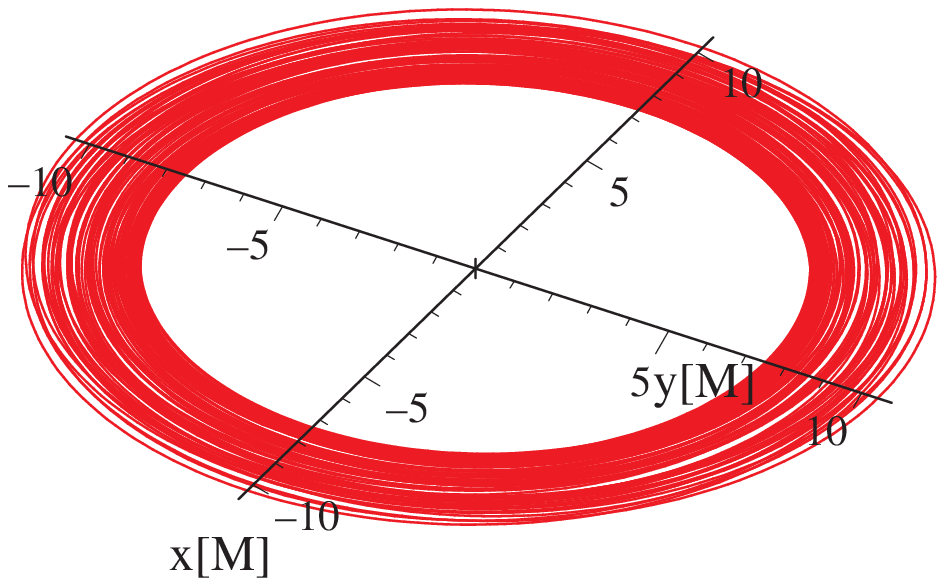}}}
\end{minipage}%
\hspace{0.5cm}
\begin{minipage}[t]{0.3 \textwidth}
\centering
\subfigure[\hspace{0.2cm} tanh ($\lm = 1$)]{
\label{fig:tanh:l=1}
\rotatebox{0}{\includegraphics[width = 5.0cm, height = 3.5cm, scale = 1]{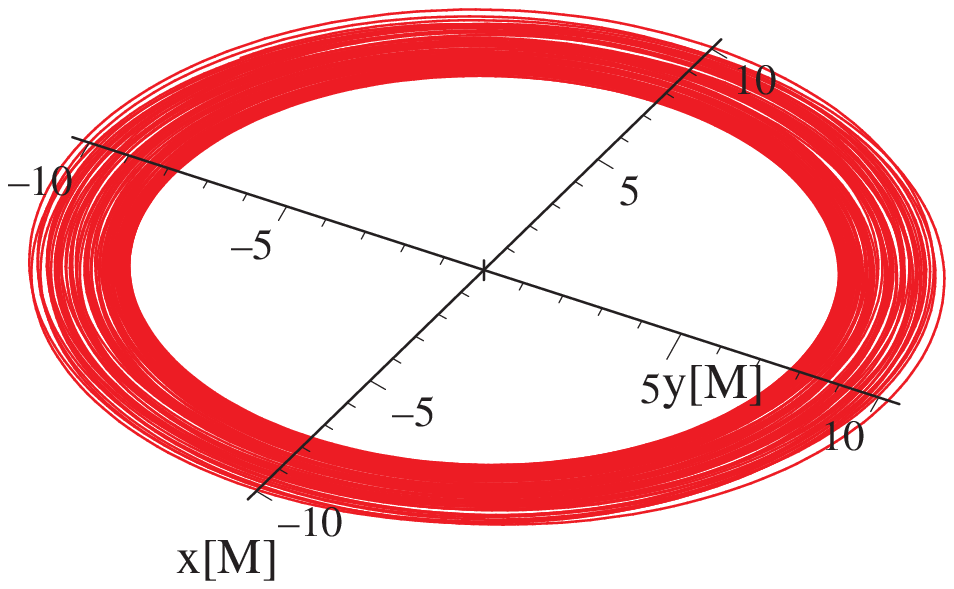}}}
\end{minipage}%
\hspace{0.5cm}
\begin{minipage}[t]{0.3 \textwidth}
\centering
\subfigure[\hspace{0.2cm} exp ($\lm = 1$)]{
\label{fig:exp:l=1}
\rotatebox{0}{\includegraphics[width = 5.0cm, height = 3.5cm, scale = 1]{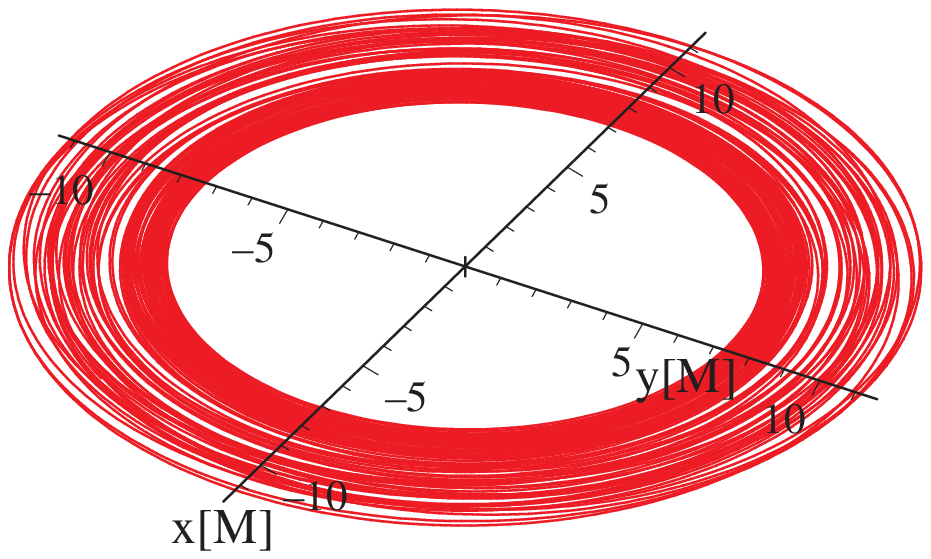}}}
\end{minipage}
\caption{\label{fig:mass-fn-orbits} Orbital trajectories for a spinning particle around accreting masses described
by different mass functions.
For all cases considered here, $\varepsilon = 0.10$ and the initial spin orientation is $\hat{\th} = \hat{\ph} = \pi/4$.
Figs.~\ref{fig:log:l=1}--\ref{fig:exp:l=1} indicate that the particle's polar angular position oscillates about
$\th = \pi/2$ due to MPD spin-gravity coupling, as observed in Fig.~\ref{fig:sch-orbit:l=1}.
\vspace{1mm}}
\end{figure*}
%

\section{Framework for Numerical Simulation}

\subsection{Equations of motion}

Before proceeding to any numerical analysis of the MPD equations (\ref{Dp=})--(\ref{Ds=}), it is useful to express
them in a purely vector form, since issues concerning roundoff errors become more manageable than otherwise.
This requires writing the equations in terms of a spin vector $S^\alpha$, defined as
\be
S^\alpha & \equiv & {1 \over 2m} \, \epsilon^\alpha{}_{\beta \mu \nu} \, P^\beta \, S^{\mu \nu},
\label{Svector=}
\ee
where $\epsilon_{\mu \nu \alpha \beta} = \sqrt{-g} \, \sigma_{\mu \nu \al \bt}$ is the Levi-Civita tensor
described in terms of the alternating symbol $\sigma_{\mu \nu \al \bt}$ with $\sigma_{0123} = 1$.
Then the spin tensor in terms of $S^\alpha$ is
\be
S^{\al \bt} & = & {1 \over m} \, \epsilon^{\al \bt \mu \nu} \, P_\mu \, S_\nu,
\label{Stensor=}
\ee
where $\epsilon^{\mu \nu \alpha \beta} = -{1 \over \sqrt{-g}} \, \sigma^{\mu \nu \al \bt}$.
The spin condition
\be
S^\al \, P_\al & = & 0
\label{s.p=0}
\ee
then naturally follows from (\ref{Svector=}), while the spin magnitude ${\cal S}$ is
defined by
\be
{\cal S}^2 & = & S^\al \, S_\al \ = \ {1 \over 2} \, S^{\al \bt} \, S_{\al \bt},
\label{s.s=}
\ee
with units of $m \, M(\nu)$.

Besides the MPD equations, the time evolution of $\nu$ must also be taken into account, since information about
the mass function $M(\nu)$ has to be related to the particle's position with respect to co-ordinate time $t$
in a consistent way.
This requires an extension of the full set of equations to incorporate the algebraic constraint given by the
tortoise co-ordinate condition (\ref{nu=}), resulting in a Type I differential-algebraic system of equations
to solve \cite{Hairer}.
Therefore, the extended MPD equations for numerical integration are
\begin{subequations}
\label{MPD-eq}
\be
{dP^\alpha \over dt} & = & - \Gamma^\alpha{}_{\mu \nu} \, P^\mu \, V^\nu
\nn
& &{} + \lambda \left[{1 \over 2m} \, R^\alpha{}_{\beta \rho \sigma} \,
\epsilon^{\rho \sigma}{}_{\mu \nu} \, S^\mu \, P^\nu \, V^\beta \right] ,
\label{dp/dt}
\nl
{dS^\alpha \over dt} & = & - \Gamma^\alpha{}_{\mu \nu} \, S^\mu \, V^\nu
\nn
& &{} + \lambda \left[{1 \over 2m^3} \, R_{\gamma \beta \rho \sigma} \,
\epsilon^{\rho \sigma}{}_{\mu \nu} \, S^\mu \, P^\nu \, S^\gamma \, V^\beta \right]P^\alpha,
\label{ds/dt}
\nl
{dx^\alpha \over dt} & = & -{P \cdot V \over m^2}
\left[P^\alpha
+ {1 \over 2} \, {\lambda \left(S^{\alpha \beta} \, R_{\beta \gamma \mu \nu} \, P^\gamma \, S^{\mu \nu} \right)
\over m^2 + {1 \over 4} \, \lambda \left(R_{\mu \nu \rho \sigma} \, S^{\mu \nu} \, S^{\rho \sigma}\right)} \right] ,
\label{dx/dt}
\nn
\nl
{d\nu \over dt} & = & \left(\partial \nu \over \partial t \right)_r + \left(\partial \nu \over \partial r \right)_t \, {dr \over dt} ,
\label{dnu/dt}
\ee
\end{subequations}
where the spin condition (\ref{s.p=0}) is assumed throughout the particle's motion,
(\ref{nu=}) is utilized to describe (\ref{dnu/dt}), and $\lambda$ is a dimensionless parameter added
to tag all terms with MPD spin-gravity coupling.
This turns out to be a very useful device, since the tuning of $\lambda$ allows for direct comparison between spin evolution
involving strictly parallel transport of the spin vector $(\lambda = 0)$ with the contribution due to
spin-gravity coupling $(\lambda = 1)$.
\begin{figure*}
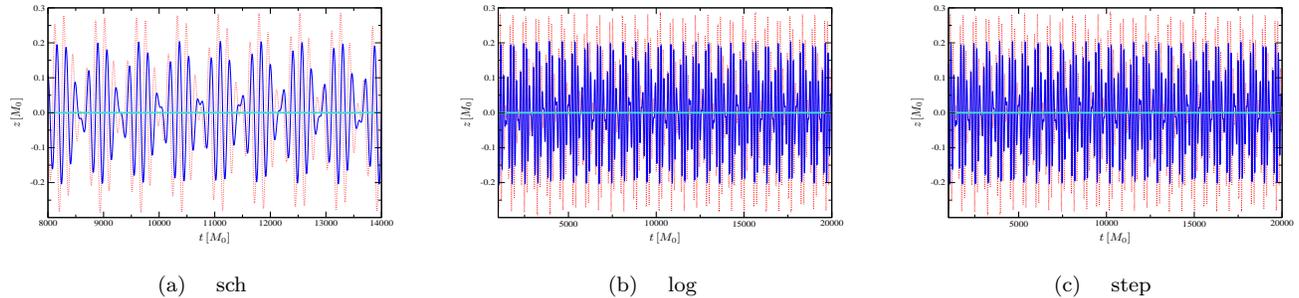

\psfrag{t [M]}[tc][][2.5][0]{$t \, [M_0]$}
\psfrag{z [M]}[bc][][2.5][0]{$z \, [M_0]$}
\vspace{1mm}
\begin{minipage}[t]{0.3 \textwidth}
\centering
\subfigure[\hspace{0.2cm} sch]{
\label{fig:z-sch:l=1-e=001}
\rotatebox{0}{\includegraphics[width = 5.0cm, height = 3.1cm, scale = 1]{z-new_sch_l=1_m=100E-2_sm=500E-2_v=350E-1_angles}}}
\end{minipage}%
\hspace{0.5cm}
\begin{minipage}[t]{0.3 \textwidth}
\centering
\subfigure[\hspace{0.2cm} log]{
\label{fig:z-log:l=1-e=001}
\rotatebox{0}{\includegraphics[width = 5.0cm, height = 3.1cm, scale = 1]{z-new_log_l=1_m=100E-2_sm=500E-2_v=350E-1_angles}}}
\end{minipage}%
\hspace{0.5cm}
\begin{minipage}[t]{0.3 \textwidth}
\centering
\subfigure[\hspace{0.2cm} step]{
\label{fig:z-stp:l=1-e=001}
\rotatebox{0}{\includegraphics[width = 5.0cm, height = 3.1cm, scale = 1]{z-new_stp_l=1_m=100E-2_sm=500E-2_v=350E-1_angles}}}
\end{minipage}
\caption{\label{fig:z} Particle displacement off the orbital plane for Schwarzschild, logarithmic, and step mass functions,
where $\varepsilon = 0.01.$  The dotted line corresponding to the largest amplitude refers to $\hat{\th} = \pi/2$ and $\hat{\ph} = 0$,
the next largest amplitude is described by the thin solid line for $\hat{\th} = \hat{\ph} = \pi/2$, and the thick solid horizontal line
corresponds to $\hat{\th} = 0$.
\vspace{1mm}}
\end{figure*}

\subsection{Numerical details}

\begin{figure}
\psfrag{t [M]}[tc][][2.5][0]{$t \, [M_0]$}
\psfrag{r [M]}[bc][][2.5][0]{$r \, [M_0]$}
\psfrag{l = 1}[cc][][1.5][0]{$\lm = 1$}
\psfrag{l = 0}[cc][][1.5][0]{$\lm = 0$}
\begin{minipage}[t]{0.3 \textwidth}
\centering
\subfigure[\hspace{0.2cm} tanh ($\hat{\th} = 0$)]{
\label{fig:radius-tan:th=0}
\rotatebox{0}{\includegraphics[width = 6.0cm, height = 4.0cm, scale = 1]{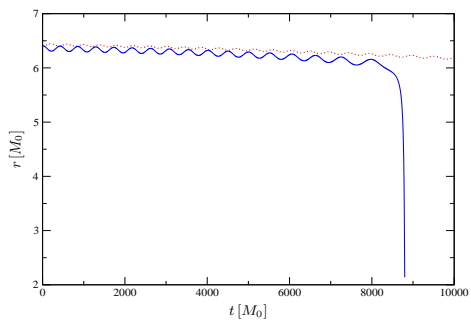}}}
\vspace{5mm}
\end{minipage}
\begin{minipage}[t]{0.3 \textwidth}
\centering
\subfigure[\hspace{0.2cm} tanh ($\hat{\th} = \pi$)]{
\label{fig:radius-tan:th=pi}
\rotatebox{0}{\includegraphics[width = 6.0cm, height = 4.0cm, scale = 1]{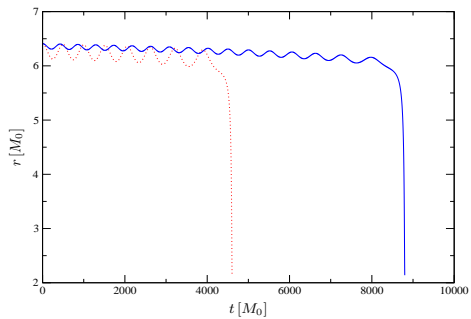}}}
\end{minipage}
\caption{\label{fig:radius-tan} Effects of initial spin orientation on the particle's orbital stability, assuming
the hyperbolic tangent mass function with $\varepsilon = 0.01$, $v = 0.395 \, c$, and ${\cal S}/m = 5 \times 10^{-3} \, M_0$.
The solid line corresponds to $\lm = 0$, while the dotted line refers to $\lm = 1$.
From Fig.~\ref{fig:radius-tan:th=0}, it is clear that spin orientation along $\hat{\th} = 0$ leads to orbital
stability when MPD spin-gravity coupling is taken into account, while from Fig.~\ref{fig:radius-tan:th=pi}
an orientation along $\hat{\th} = \pi$ leads to an accelerated orbital inspiral.}
\end{figure}
The integration is performed in Fortran 90, using a Runge-Kutta fourth-order algorithm with fixed time steps.
For each simulation run presented here, the particle is initially positioned in the $x$-direction of the
co-ordinate frame defined by the central mass function located at the origin, and follows a counterclockwise
orbit.
This is illustrated in Figure~\ref{fig:spin-orientation}, where
the particle's initial spin orientation is defined by $(\hat{\th}, \hat{\ph})$ co-ordinates following the
standard definition of spherical co-ordinates with respect to the particle's local frame.

The central mass for the Vaidya metric can be described by an arbitrary function of $\nu$, so long as
it is monotonically increasing to preserve the weak energy condition.
With this in mind, a variety of mass functions are considered in the hope of establishing a correspondence
between the time development of $M(\nu)$ and the dynamical response of the spinning point particle.
For the purpose of making well-defined comparisons, the mass functions are predefined to remain constant
from the start of the simulation until $\nu = \nu_0$, when the mass accretion rate is then ``switched on'' and
$M(\nu)$ is allowed to increase by a small relative value $\varepsilon < 1$ over some predetermined lifetime $\tau_0$
before it gets ``switched off''.
While there is no claim that this construction for the mass functions accurately reflects physically realistic scenarios,
the point of this exercise is to demonstrate the robustness of the MPD equations to
describe the effects of the particle's spin on its orbital dynamics, depending on the growth rate of the
mass increase.

All dimensional parameters are described in terms of $M_0$, the mass of the central body prior to the start of
mass accretion at $\nu = \nu_0$.
In units of $M_0$, the mass functions considered are the following:
\begin{subequations}
\label{mass-functions}
\be
M(\nu) & = & 1,
\label{sch}
\nl
& = & 1 + \varepsilon \, \ln[1 + (\nu - \nu_0)/\tau_0],
\label{log}
\nl
& = & 1 + \varepsilon \, \left[\exp\left[(\nu - \nu_0)/\tau_0\right]- 1\right],
\label{exp}
\nl
& = & 1 + {\varepsilon \over 2} \, \left\{\tanh\left[(\nu - \nu_0)/\tau_0\right] + 1\right\},
\label{tanh}
\ee
\end{subequations}
where (\ref{sch}) corresponds to a Schwarzschild black hole, (\ref{log}) describes a logarithmic
mass increase, (\ref{exp}) refers to an exponential increase, and (\ref{tanh}) defines an increase
due to a hyperbolic tangent function.
For future reference, these functions are respectively denoted by {\bf sch}, {\bf log}, {\bf exp}, and {\bf tanh} in the
plots presented in this paper.
All simulations presented for these functions assume $\tau_0 = 10^4 \, M_0$, and the choice of $\nu_0$ for each mass function
is such that $M(\nu) = 1 + \varepsilon/2$ at around $t = 10^4 \, M_0$.
The relatively large choice for $\tau_0$ is to allow for the mass functions to gradually increase over several orbital cycles.
In addition to these mass functions, a step function mass increase (denoted hereafter by {\bf step}) is
simulated by using (\ref{tanh}) with $\tau_0 = 1 \, M_0$.
A plot of the mass functions over time is presented in Figure~\ref{fig:mass-functions}.

\section{Numerical Analysis}

With the equations of motion (\ref{dp/dt})--(\ref{dnu/dt}) and mass functions (\ref{sch})--(\ref{tanh}), it is
possible to determine a wide variety of kinematic and dynamical properties associated with the spinning point particle.
This section details the numerical results that are obtained from this analysis.
All simulations considered here assume for initial conditions a particle mass of $m = 10^{-2} \, M_0$.
This is consistent with the assumption that the spinning particle
is very small compared to the central body in order to satisfy the test particle assumption.
The spin-to-mass ratio considered here, which also satisfies the required approximation ${\cal S}/(m \, r) \ll 1$,
is ${\cal S}/m = 5 \times 10^{-2} \, M_0$, with exceptions noted as required.
As with the choice of $m$, the spin magnitude is sufficiently large to accommodate for realistic astrophysical conditions
while avoiding numerical instabilities that follow from making ${\cal S}$ too large.
Except where otherwise indicated, 
$\varepsilon = 0.01$ and the initial orbital speed of the particle is $v = 0.350 \, c$,
which corresponds to an initial radius of $r \approx 8 \, M_0$.


\subsection{Orbital kinematics}

\begin{figure}
\psfrag{t [M]}[tc][][2.5][0]{$t \, [M_0]$}
\psfrag{theta (radians)}[bc][][2.5][0]{$\th$ (radians)}
\psfrag{l = 1}[cc][][1.5][0]{$\lm = 1$}
\psfrag{l = 0}[cc][][1.5][0]{$\lm = 0$}
\begin{minipage}[t]{0.3 \textwidth}
\centering
\subfigure[\hspace{0.2cm} log ($\hat{\th} = \hat{\ph} = \pi/4$)]{
\label{fig:theta-log}
\rotatebox{0}{\includegraphics[width = 6.0cm, height = 4.0cm, scale = 1]{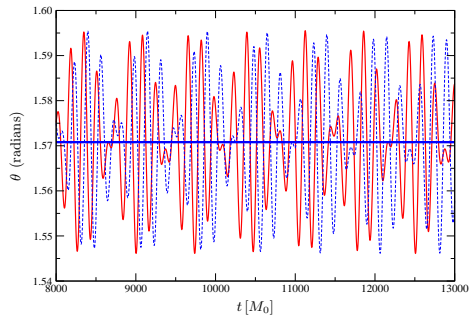}}}
\vspace{5mm}
\end{minipage}
\begin{minipage}[t]{0.3 \textwidth}
\centering
\subfigure[\hspace{0.2cm} step ($\hat{\th} = \hat{\ph} = \pi/4$)]{
\label{fig:theta-stp}
\rotatebox{0}{\includegraphics[width = 6.0cm, height = 4.0cm, scale = 1]{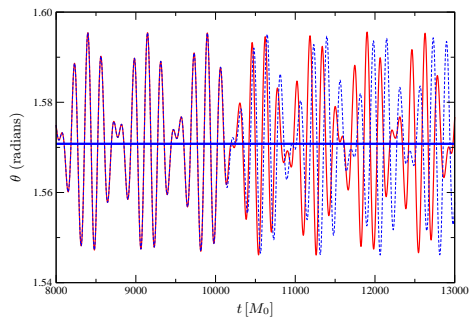}}}
\end{minipage}
\caption{\label{fig:theta} Polar angular displacement for Schwarzschild, logarithmic, and step mass functions, where
$\varepsilon = 0.01$ and $\hat{\th} = \hat{\ph} = \pi/4$.
The oscillating solid and dashed lines correspond to $\lm = 1$, where the latter describes the displacement due to
a Schwarzschild black hole.
In contrast, the solid horizontal line for $\lm = 0$ shows that MPD spin-gravity coupling is the sole contributor to the
polar displacement observed.}
\end{figure}
\begin{figure}
\psfrag{t [M]}[tc][][2.5][0]{$t \, [M_0]$}
\psfrag{phi (radians)}[bc][][2.5][0]{$\ph$ (radians)}
\psfrag{l = 1}[cc][][1.5][0]{$\lm = 1$}
\psfrag{l = 0}[cc][][1.5][0]{$\lm = 0$}
\begin{minipage}[t]{0.3 \textwidth}
\centering
\subfigure[\hspace{0.2cm} sch ($\hat{\th} = \hat{\ph} = \pi/4$)]{
\label{fig:phi-sch}
\rotatebox{0}{\includegraphics[width = 6.0cm, height = 4.0cm, scale = 1]{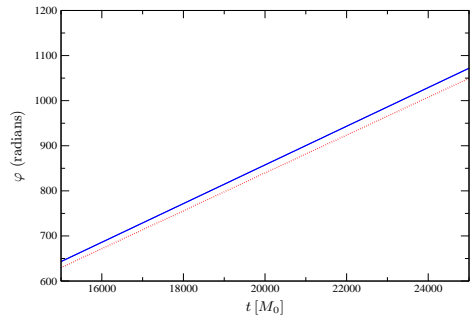}}}
\vspace{5mm}
\end{minipage}
\begin{minipage}[t]{0.3 \textwidth}
\centering
\subfigure[\hspace{0.2cm} tanh]{
\label{fig:phi-tan:th=0}
\rotatebox{0}{\includegraphics[width = 6.0cm, height = 4.0cm, scale = 1]{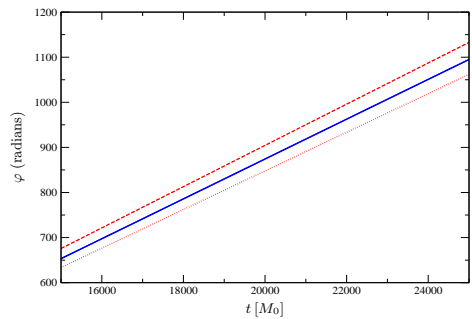}}}
\end{minipage}
\caption{\label{fig:phi} Azimuthal angular displacement for Schwarzschild and hyperbolic tangent mass functions, where
$\varepsilon = 0.01$.
For Fig.~\ref{fig:phi-tan:th=0}, the effect of MPD spin-gravity coupling leads to a decrease in angular displacement
compared to geodesic motion ($\lm = 0$) due to $\hat{\th} = 0$ spin orientation, as described by the dotted line.
The dashed line corresponding to $\hat{\th} = \pi$ shows a slight increase by roughly the same amount.}
\end{figure}
A sampling of orbital trajectories for a classical spinning particle are presented in Figures~\ref{fig:sch-orbit}
and~\ref{fig:mass-fn-orbits}, where Figure~\ref{fig:sch-orbit} describes the orbit around a Schwarzschild black hole
and Figure~\ref{fig:mass-fn-orbits} lists the orbits for the logarithmic, hyperbolic tangent, and exponential mass functions,
where $\varepsilon = 0.10$.
As well, the initial spin orientation is $\hat{\th} = \hat{\ph} = \pi/4$ for each plot.

From Figure~\ref{fig:sch-orbit:l=0}, it is clear that the particle follows a stable circular orbit due to purely geodesic
motion $(\lm~=~0)$, while Figure~\ref{fig:sch-orbit:l=1} shows the contribution of the MPD spin-gravity coupling $(\lm~=~1)$
with a slight quasiperiodic oscillation of the particle's position about the $\th = \pi/2$ orbital plane.
However, for these initial conditions the orbital radius remains largely unchanged due to the force and torque terms
in the MPD equations.
In contrast, Figure~\ref{fig:mass-fn-orbits} shows that the accreting mass can significantly alter the trajectory for sufficiently
large $\varepsilon$, leading to precessing ellipsoidal orbits.
In particular, it is clear from comparing Figures~\ref{fig:log:l=0}--\ref{fig:exp:l=0} to Figures~\ref{fig:log:l=1}--\ref{fig:exp:l=1}
that the influence of MPD spin-gravity coupling is relatively small compared to the particle's centre-of-mass motion due to the
geodesic equation.
Nonetheless, there are differences in that there is both some oscillation about the orbital plane and
a slight reduction of the orbital eccentricity due to the presence of spin.
Not surprisingly, the orbiting particle responds in proportion to the degree of central mass increase with time, and that the
MPD force and torque terms are similarly responsive for the assumed spin orientation.

\begin{figure*}
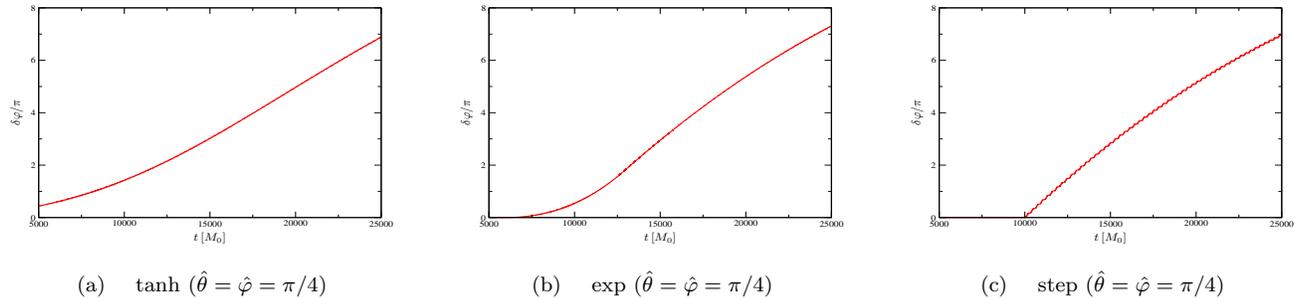

\psfrag{t [M]}[tc][][2.5][0]{$t \, [M_0]$}
\psfrag{d_phi/Pi}[bc][][2.5][0]{$\delta \ph/\pi$}
\vspace{1mm}
\begin{minipage}[t]{0.3 \textwidth}
\centering
\subfigure[\hspace{0.2cm} tanh ($\hat{\th} = \hat{\ph} = \pi/4$)]{
\label{fig:phase-tan:l=1-e=001}
\rotatebox{0}{\includegraphics[width = 5.0cm, height = 3.1cm, scale = 1]{phase-new_tan_l=1_m=100E-2_sm=500E-2_v=350E-1_theta=025_phi=025}}}
\end{minipage}%
\hspace{0.5cm}
\begin{minipage}[t]{0.3 \textwidth}
\centering
\subfigure[\hspace{0.2cm} exp ($\hat{\th} = \hat{\ph} = \pi/4$)]{
\label{fig:phase-exp:l=1-e=001}
\rotatebox{0}{\includegraphics[width = 5.0cm, height = 3.1cm, scale = 1]{phase-new_exp_l=1_m=100E-2_sm=500E-2_v=350E-1_theta=025_phi=025}}}
\end{minipage}%
\hspace{0.5cm}
\begin{minipage}[t]{0.3 \textwidth}
\centering
\subfigure[\hspace{0.2cm} step ($\hat{\th} = \hat{\ph} = \pi/4$)]{
\label{fig:phase-stp:l=1-e=001}
\rotatebox{0}{\includegraphics[width = 5.0cm, height = 3.1cm, scale = 1]{phase-new_stp_l=1_m=100E-2_sm=500E-2_v=350E-1_theta=025_phi=025}}}
\end{minipage}
\caption{\label{fig:phase} Phase difference in particle's centre-of-mass azimuthal position due to dynamical central mass
when compared to Schwarzschild mass for $\lm = 1$.
Figs.~\ref{fig:phase-tan:l=1-e=001}--\ref{fig:phase-stp:l=1-e=001} show that mass accretion induces a phase advance in the
particle's orbit that reflects the rate of mass increase.
\vspace{1mm}}
\end{figure*}
To determine the effect of MPD spin-gravity coupling on particle confinement to the orbital plane, Figure~\ref{fig:z}
describes its $z$-component amplitude over time for the Schwarzschild, logarithmic, and step mass functions.
For spin orientation prepared along the radial direction $(\hat{\th} = \pi/2$ and $\hat{\ph} = 0)$, the amplitude
is $z \approx 0.3 \, M_0$, while an orientation along the tangential direction $(\hat{\th} = \hat{\ph} = \pi/2)$
produces a slightly reduced amplitude of $z \approx 0.2 \, M_0$.
This leads to a 2.50-3.75\% displacement when compared to the particle's initial radial position.
It is also clear from Figures~\ref{fig:z-sch:l=1-e=001}--\ref{fig:z-stp:l=1-e=001} that a spin orientation normal to
the orbital plane $(\hat{\th} = 0)$ produces no obvious displacement.
In addition, there is no evidence of amplitude growth due to mass accretion, regardless of the particle's spin orientation.

It is of interest to examine the influence of spin on the particle's orbital stability over time for situations
where the centre-of-mass motion alone leads to an inspiral.
This is demonstrated in Figure~\ref{fig:radius-tan} for the hyperbolic tangent mass function,
which compares the lifetime of the particle's orbit, where $v = 0.395 \, c$.
Comparison of Figures~\ref{fig:radius-tan:th=0} and \ref{fig:radius-tan:th=pi} shows that the spin component along the
$\th = 0$ direction is solely responsible for producing stability in the particle's orbit, and that a spin component directed
along the $\th = \pi$ direction increases the inspiral rate when compared to purely geodesic motion.
Given that most (if not all) observed binary systems have their orbital and spin angular momenta pointed
in the same general direction, this result makes intuitive sense.

As demonstrated by the orbital trajectories of Figures~\ref{fig:sch-orbit} and~\ref{fig:mass-fn-orbits}, there is an oscillation
of the particle's polar angular position about $\th = \pi/2$ due to the force and torque terms in the MPD equations of motion.
A closer examination of this effect is presented in Figure~\ref{fig:theta}, with a comparison of the polar angle displacement for
Schwarzschild, logarithmic, and step mass functions as a function of time, with $\varepsilon = 0.01$.
All plots here show that a polar displacement of $\Delta \th \approx 7.7 \times 10^{-3} \, \pi$ about the $\th = \pi/2$ orbital plane
occurs due to MPD spin-gravity coupling.
When mass accretion is taken into account, the frequency of oscillation increases in accordance with the rate of mass increase,
as suggested by Figures~\ref{fig:theta-log} and~\ref{fig:theta-stp}, though the maximum displacement from $\th = \pi/2$ remains largely
constant.

\begin{figure}
\psfrag{t [M]}[tc][][2.5][0]{$t \, [M_0]$}
\psfrag{P_theta}[cc][][2.5][0]{$P^\th$}
\psfrag{l = 1}[cc][][1.5][0]{$\lm = 1$}
\psfrag{l = 0}[cc][][1.5][0]{$\lm = 0$}
\begin{minipage}[t]{0.3 \textwidth}
\centering
\subfigure[\hspace{0.2cm} log ($\hat{\th} = \hat{\ph} = \pi/4$)]{
\label{fig:P2-log:th=ph=pi/4}
\rotatebox{0}{\includegraphics[width = 6.0cm, height = 4.0cm, scale = 1]{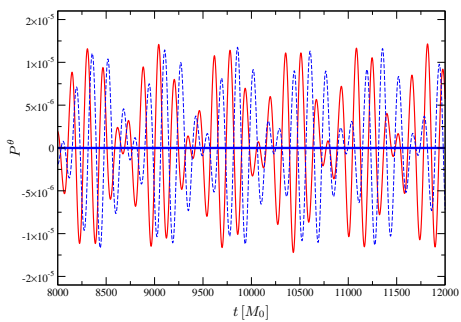}}}
\end{minipage}%
\vspace{7mm}
\begin{minipage}[t]{0.3 \textwidth}
\centering
\subfigure[\hspace{0.2cm} tanh ($\hat{\th} = \hat{\ph} = \pi/4$)]{
\label{fig:P2-tan:th=ph=pi/4}
\rotatebox{0}{\includegraphics[width = 6.0cm, height = 4.0cm, scale = 1]{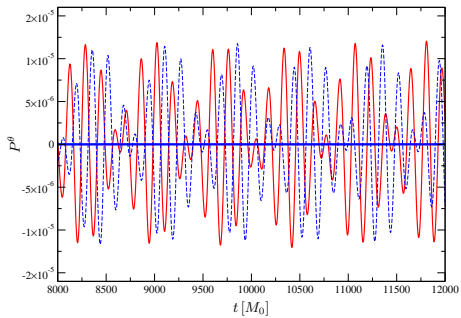}}}
\end{minipage}
\caption{\label{fig:P2} Polar component of the particle's four-momentum for the logarithmic and hyperbolic tangent
mass functions, where $\varepsilon = 0.01$.
The dashed line describes the amplitude due to a Schwarzschild black hole.
\vspace{1mm}}
\end{figure}
\begin{figure*}
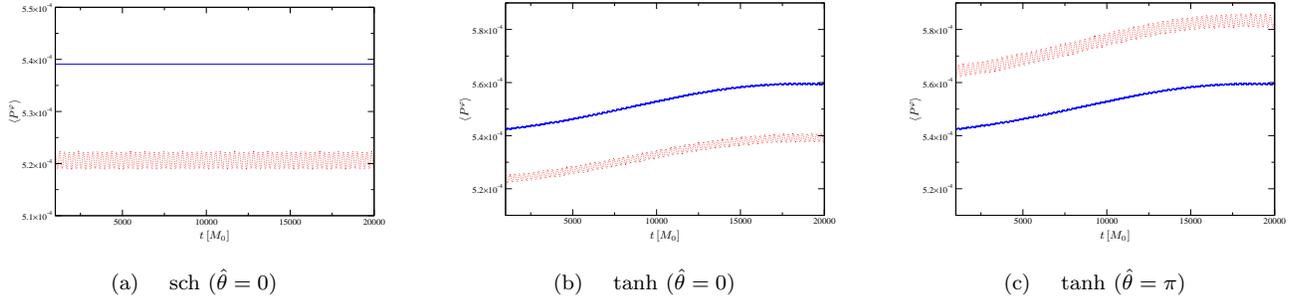

\psfrag{t [M]}[tc][][2.5][0]{$t \, [M_0]$}
\psfrag{<P_phi>}[cc][][2.5][0]{$\langle P^\ph \rangle$}
\psfrag{l = 1}[cc][][1.5][0]{$\lm = 1$}
\psfrag{l = 0}[cc][][1.5][0]{$\lm = 0$}
\vspace{1mm}
\begin{minipage}[t]{0.3 \textwidth}
\centering
\subfigure[\hspace{0.2cm} sch ($\hat{\th} = 0$)]{
\label{fig:P3-sch:th=0}
\rotatebox{0}{\includegraphics[width = 5.0cm, height = 3.1cm, scale = 1]{P3-new_sch_m=100E-2_sm=500E-2_v=350E-1_theta=000_phi=000}}}
\end{minipage}%
\hspace{0.5cm}
\begin{minipage}[t]{0.3 \textwidth}
\centering
\subfigure[\hspace{0.2cm} tanh ($\hat{\th} = 0$)]{
\label{fig:P3-tan:th=0}
\rotatebox{0}{\includegraphics[width = 5.0cm, height = 3.1cm, scale = 1]{P3-new_tan_m=100E-2_sm=500E-2_v=350E-1_theta=000_phi=000}}}
\end{minipage}%
\hspace{0.5cm}
\begin{minipage}[t]{0.3 \textwidth}
\centering
\subfigure[\hspace{0.2cm} tanh ($\hat{\th} = \pi$)]{
\label{fig:P3-tan:th=pi}
\rotatebox{0}{\includegraphics[width = 5.0cm, height = 3.1cm, scale = 1]{P3-new_tan_m=100E-2_sm=500E-2_v=350E-1_theta=100_phi=000}}}
\end{minipage}
\caption{\label{fig:P3} Azimuthal component of the particle's four-momentum for Schwarzschild and hyperbolic tangent
mass functions, where $\varepsilon = 0.01$.
Fig.~\ref{fig:P3-tan:th=0} shows that the MPD spin-gravity coupling reduces the orbital speed for $\hat{\th} = 0$,
while the opposite is true for $\hat{\th} = \pi$, as shown by Fig.~\ref{fig:P3-tan:th=pi}.
\vspace{1mm}}
\end{figure*}
It is also evident from Figure~\ref{fig:phi} that the MPD spin-gravity coupling has an influence on the overall angular displacement in
the azimuthal direction, depending on the initial spin orientation.
For spin orientation with a component along $\th = 0$, the particle's orbital frequency decreases when compared to that due to
centre-of-mass motion only, while a component along $\th = \pi$ drives up the frequency by about the same rate.
This is demonstrated by Figure~\ref{fig:phi-tan:th=0} for the hyperbolic tangent mass function.
There is no observed difference in the particle's azimuthal position due to MPD spin-gravity coupling for spin orientations confined to
the orbital plane.

Another interesting feature is presented in Figure~\ref{fig:phase}, which plots the degree of phase advance defined by
\be
\delta \ph(t) & \equiv & \ph(t)_{\rm Vaidya} - \ph(t)_{\rm Sch}
\label{phase-advance}
\ee
for a given moment in time during the simulation.
Comparing Figures~\ref{fig:phase-tan:l=1-e=001}--\ref{fig:phase-stp:l=1-e=001} for the hyperbolic tangent, exponential, and
step mass functions, respectively, it appears that the amount of phase advance over time is roughly proportional to the rate
of mass increase at a given moment.
Given that the orbital period is around $146.5 \, M_0$, a $\pi/2$ phase shift beyond what is plotted at $t = 10^4 \, M_0$ for the
hyperbolic tangent function occurs after about 14 particle orbits, while the same effect occurs after roughly 10 orbits for
the exponential function.
Reflecting the sudden mass increase at $t = 10^4 \, M_0$ for the step mass function, the $\pi/2$ shift occurs in just five
orbits of the particle.

\subsection{Particle dynamics}

\begin{figure*}
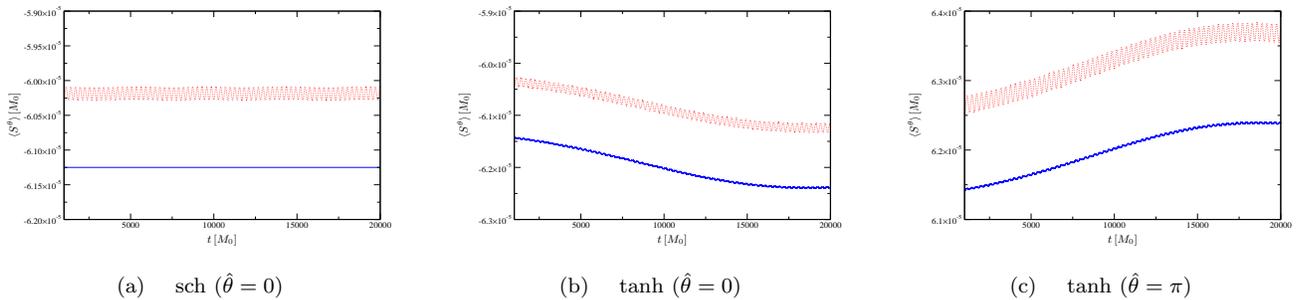

\psfrag{t [M]}[tc][][2.5][0]{$t \, [M_0]$}
\psfrag{<S_theta> [M]}[bc][][2.5][0]{$\langle S^\th \rangle \, [M_0]$}
\psfrag{l = 1}[cc][][1.5][0]{$\lm = 1$}
\psfrag{l = 0}[cc][][1.5][0]{$\lm = 0$}
\vspace{1mm}
\begin{minipage}[t]{0.3 \textwidth}
\centering
\subfigure[\hspace{0.2cm} sch ($\hat{\th} = 0$)]{
\label{fig:S2-sch:th=0}
\rotatebox{0}{\includegraphics[width = 5.0cm, height = 3.1cm, scale = 1]{S2-new_sch_m=100E-2_sm=500E-2_v=350E-1_theta=000_phi=000}}}
\end{minipage}%
\hspace{0.5cm}
\begin{minipage}[t]{0.3 \textwidth}
\centering
\subfigure[\hspace{0.2cm} tanh ($\hat{\th} = 0$)]{
\label{fig:S2-tan:th=0}
\rotatebox{0}{\includegraphics[width = 5.0cm, height = 3.1cm, scale = 1]{S2-new_tan_m=100E-2_sm=500E-2_v=350E-1_theta=000_phi=000}}}
\end{minipage}%
\hspace{0.5cm}
\begin{minipage}[t]{0.3 \textwidth}
\centering
\subfigure[\hspace{0.2cm} tanh ($\hat{\th} = \pi$)]{
\label{fig:S2-tan:th=pi}
\rotatebox{0}{\includegraphics[width = 5.0cm, height = 3.1cm, scale = 1]{S2-new_tan_m=100E-2_sm=500E-2_v=350E-1_theta=100_phi=000}}}
\end{minipage}
\caption{\label{fig:S2} Polar component of the particle's spin angular momentum for Schwarzschild and hyperbolic tangent
mass functions, where $\varepsilon = 0.01$.
The mass accretion moderately boosts the magnitude of $S^\theta$, as shown by comparing
Figs.~\ref{fig:S2-sch:th=0}--\ref{fig:S2-tan:th=0}.
The MPD spin-gravity coupling reduces the spin magnitude for $\hat{\th} = 0$,
while Fig.~\ref{fig:S2-tan:th=pi} shows an increase in the magnitude for $\hat{\th} = \pi$.
\vspace{1mm}}
\end{figure*}
\begin{figure*}
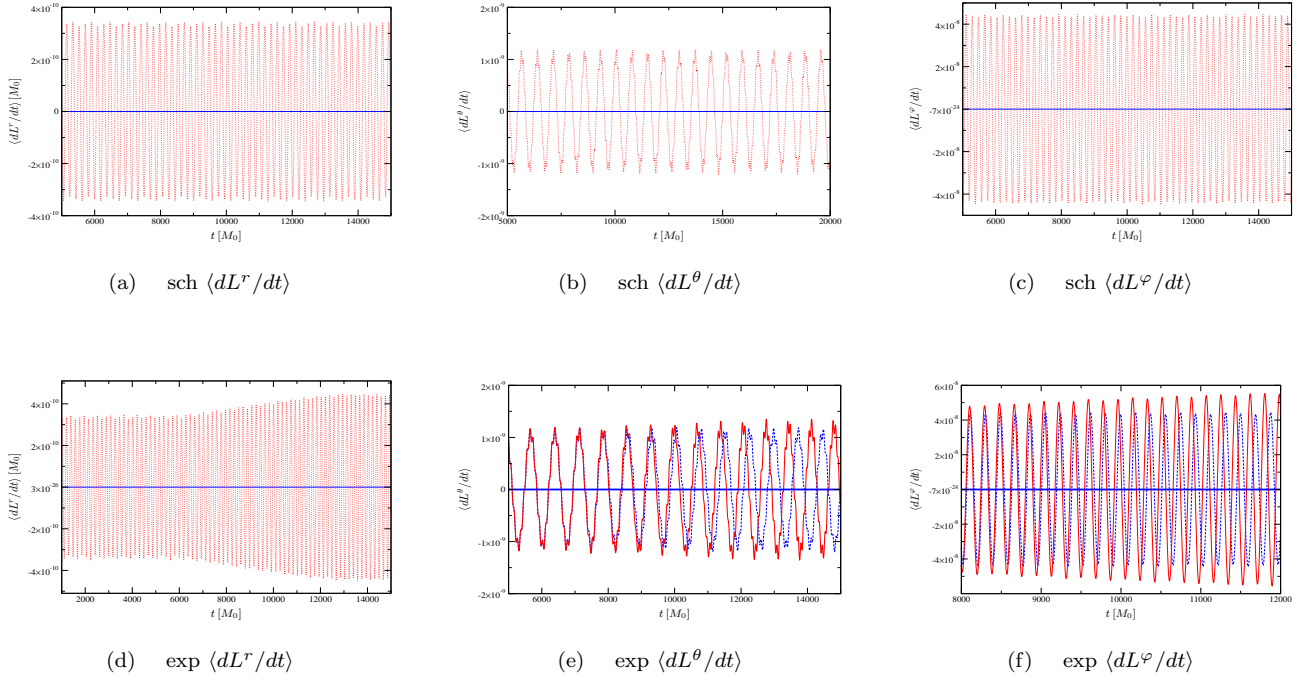

\psfrag{t [M]}[tc][][2.5][0]{$t \, [M_0]$}
\psfrag{<dL_r/dt> [M]}[bc][][2.5][0]{$\langle dL^r/dt \rangle \, [M_0]$}
\psfrag{<dL_theta/dt>}[bc][][2.5][0]{$\langle dL^\th/dt \rangle$}
\psfrag{<dL_phi/dt>}[bc][][2.5][0]{$\langle dL^\ph/dt \rangle$}
\psfrag{l = 1}[cc][][1.5][0]{$\lm = 1$}
\psfrag{l = 0}[cc][][1.5][0]{$\lm = 0$}
\vspace{1mm}
\begin{minipage}[t]{0.3 \textwidth}
\centering
\subfigure[\hspace{0.2cm} sch $\langle dL^r/dt \rangle$]{
\label{fig:torque-r-sch}
\rotatebox{0}{\includegraphics[width = 5.0cm, height = 3.1cm, scale = 1]{torque_r-new_sch_m=100E-2_sm=500E-2_v=350E-1_theta=025_phi=025}}}
\end{minipage}%
\hspace{0.5cm}
\begin{minipage}[t]{0.3 \textwidth}
\centering
\subfigure[\hspace{0.2cm} sch $\langle dL^\th/dt \rangle$]{
\label{fig:torque-t-sch}
\rotatebox{0}{\includegraphics[width = 5.0cm, height = 3.1cm, scale = 1]{torque_t-new_sch_m=100E-2_sm=500E-2_v=350E-1_theta=025_phi=025}}}
\end{minipage}%
\hspace{0.5cm}
\begin{minipage}[t]{0.3 \textwidth}
\centering
\subfigure[\hspace{0.2cm} sch $\langle dL^\ph/dt \rangle$]{
\label{fig:torque-p-sch}
\rotatebox{0}{\includegraphics[width = 5.0cm, height = 3.1cm, scale = 1]{torque_p-new_sch_m=100E-2_sm=500E-2_v=350E-1_theta=025_phi=025}}}
\end{minipage}
\begin{minipage}[t]{0.3 \textwidth}
\vspace{4mm}
\centering
\subfigure[\hspace{0.2cm} exp $\langle dL^r/dt \rangle$]{
\label{fig:torque-r-exp:e=001}
\rotatebox{0}{\includegraphics[width = 5.0cm, height = 3.1cm, scale = 1]{torque_r-new_exp_m=100E-2_sm=500E-2_v=350E-1_theta=025_phi=025}}}
\end{minipage}%
\hspace{0.5cm}
\begin{minipage}[t]{0.3 \textwidth}
\vspace{4mm}
\centering
\subfigure[\hspace{0.2cm} exp $\langle dL^\th/dt \rangle$]{
\label{fig:torque-t-exp:e=001}
\rotatebox{0}{\includegraphics[width = 5.0cm, height = 3.1cm, scale = 1]{torque_t-new_exp_m=100E-2_sm=500E-2_v=350E-1_theta=025_phi=025}}}
\end{minipage}%
\hspace{0.5cm}
\begin{minipage}[t]{0.3 \textwidth}
\vspace{4mm}
\centering
\subfigure[\hspace{0.2cm} exp $\langle dL^\ph/dt \rangle$]{
\label{fig:torque-p-exp:e=001}
\rotatebox{0}{\includegraphics[width = 5.0cm, height = 3.1cm, scale = 1]{torque_p-new_exp_m=100E-2_sm=500E-2_v=350E-1_theta=025_phi=025}}}
\end{minipage}
\caption{\label{fig:torque:e=001} Time-averaged orbital torque on the spinning particle for Schwarzschild and exponential mass functions,
where $\varepsilon = 0.01$ and $\hat{\th} = \hat{\ph} = \pi/4$.
While some variation exists due to MPD spin-gravity coupling, Figs.~\ref{fig:torque-r-sch}--\ref{fig:torque-p-exp:e=001} show
that orbital angular momentum is still conserved.
\vspace{1mm}}
\end{figure*}
The extended MPD equations of motion (\ref{MPD-eq}) form a twelve-dimensional dynamical system for the
time evolution of the particle's position, linear momentum, and spin angular momentum vectors.
It follows that a complex interplay should exist between the linear and spin angular momentum components that leads
to the orbital kinematics described earlier.
While it is difficult to determine precisely how the interplay unfolds, it is possible to
see some interesting properties of the dynamical variables in isolation, particularly when there is a time-evolving metric
due to mass accretion.

One example is found in Figure~\ref{fig:P2}, which describes the polar component of the particle's four-momentum over time.
When mass accretion is taken into account, as shown by Figures~\ref{fig:P2-log:th=ph=pi/4}--\ref{fig:P2-tan:th=ph=pi/4},
the polar momentum component oscillates more rapidly in response to where the mass accretion rate is strongest.
A second example of note shows the time-averaged evolution of the particle's azimuthal component of its four-momentum,
described by Figure~\ref{fig:P3}.
(All time-averaged plots presented in this paper are calculated over an interval of $\Delta t = 10^3 \, M_0$.)
From Figures~\ref{fig:P3-tan:th=0} and~\ref{fig:P3-tan:th=pi}, it is evident that the MPD spin-gravity coupling acts
to reduce the tangential momentum if the spin component is projected along $\hat{\th} = 0$, while it increases $\langle P^\ph \rangle$
if the spin has a projection along $\hat{\th} = \pi$.
Similarly, from Figure~\ref{fig:S2}, the polar component of the particle's spin angular momentum exhibits a similar behaviour.
Comparing Figures~\ref{fig:S2-tan:th=0} and \ref{fig:S2-tan:th=pi}, it follows that the MPD spin-gravity coupling
reduces the magnitude of $\langle S^\th \rangle$ for projections along $\hat{\th} = 0$ while it increases the magnitude for
projections of the spin orientation in the opposite direction.
The effect of the mass accretion, as especially shown in Figures~\ref{fig:P3} and~\ref{fig:S2}, is to moderately increase
the overall magnitude of the particle's linear and spin angular momenta.

\begin{figure*}
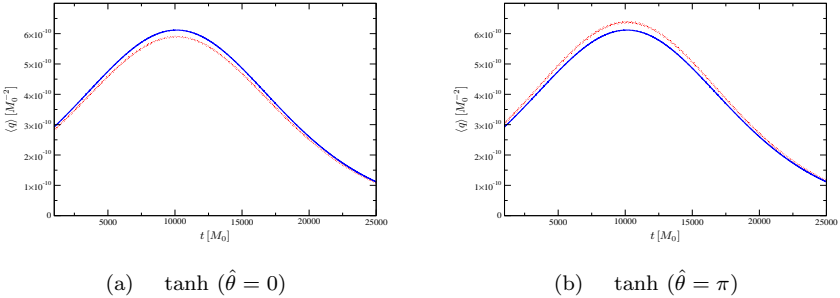

\psfrag{t [M]}[tc][][2.5][0]{$t \, [M_0]$}
\psfrag{<q> [M^(-2)]}[bc][][2.5][0]{$\langle q \rangle \, [M_0^{-2}]$}
\psfrag{l = 1}[cc][][1.5][0]{$\lm = 1$}
\psfrag{l = 0}[cc][][1.5][0]{$\lm = 0$}
\vspace{2mm}
\begin{minipage}[t]{0.3 \textwidth}
\centering
\subfigure[\hspace{0.2cm} tanh ($\hat{\th} = 0$)]{
\label{fig:rad-density:th=0}
\rotatebox{0}{\includegraphics[width = 5.0cm, height = 3.1cm, scale = 1]{rad-new_tan_m=100E-2_sm=500E-2_v=350E-1_theta=000_phi=000}}}
\end{minipage}%
\hspace{0.5cm}
\begin{minipage}[t]{0.3 \textwidth}
\centering
\subfigure[\hspace{0.2cm} tanh ($\hat{\th} = \pi$)]{
\label{fig:rad-density:th=pi}
\rotatebox{0}{\includegraphics[width = 5.0cm, height = 3.1cm, scale = 1]{rad-new_tan_m=100E-2_sm=500E-2_v=350E-1_theta=100_phi=000}}}
\end{minipage}%
\hspace{0.5cm}
\begin{minipage}[t]{0.3 \textwidth}
\centering
\vspace{-3.0cm}
\caption{\label{fig:rad-density} Radiation density for an observer comoving with the spinning particle due to
the hyperbolic tangent mass function, where $\varepsilon = 0.01$.
}
\end{minipage}
\vspace{1mm}
\end{figure*}
It is particularly interesting to investigate the effects of mass accretion on the orbital angular momentum's rate of change
with respect to time.
A straightforward calculation shows that the spinning particle's orbital angular momentum components are
\begin{subequations}
\label{L-orbit}
\be
L^r & = & 0,
\label{L-orbit-r}
\nl
L^\th & = & -r \, \sin \th \, P^\ph,
\label{L-orbit-theta}
\nl
L^\ph & = & r \, P^\th /\sin \th,
\label{L-orbit-phi}
\ee
\end{subequations}
while the components of the torque on the particle are
\begin{subequations}
\label{torque}
\be
{dL^r \over dt} & = & r^2 \, \sin \th \lt({d \th \over dt} \, P^\ph - {d \ph \over dt} \, P^\th\rt),
\label{torque-r}
\nl
{dL^\th \over dt} & = & -\lt[r \lt(\sin \th \, {d P^\ph \over dt} + \cos \th \lt({d \th \over dt} \, P^\ph
+ {d \ph \over dt} \, P^\th \rt) \rt) \rt.
\nn
& &{} + \lt. 2 \, \sin \th \, {dr \over dt} \, P^\ph \rt],
\label{torque-theta}
\nl
{dL^\ph \over dt} & = & {1 \over \sin \th} \lt[r \lt({dP^\th \over dt} - \sin \th \, \cos \th \, {d \ph \over dt} \, P^\ph \rt) \rt.
\nn
& &{} + \lt. 2 \, {dr \over dt} \, P^\th\rt].
\label{torque-phi}
\ee
\end{subequations}
For a particle in circular motion confined to the $\th =~\pi/2$ orbital plane, (\ref{torque}) is identically zero, as expected.
Figure~\ref{fig:torque:e=001} is a time-averaged plot of (\ref{torque-r})--(\ref{torque-phi}) with respect to time for the
Schwarzschild and exponential mass functions, where  $\hat{\th} = \hat{\ph} = \pi/4$.
While there is some variation present in $\langle dL^i/dt\rangle$ due to MPD spin-gravity coupling, it is quite
clear that orbital angular momentum is still a conserved quantity on average.

\subsection{Radiation density of the Vaidya Space-time background}

Because the Vaidya metric satisfies the Einstein field equations in matter described by a null fluid,
a particle in orbit will come in contact with the surrounding radiation, with possible dynamical consequences.
The radiation density corresponding to the Vaidya background, as measured by a local observer \cite{Vaidya}, is
\be
q & = & T_{\mu \nu} \, V^\mu \, V^\nu \ = \ {1 \over 8 \, \pi} \, R_{\mu \nu} \, V^\mu \, V^\nu,
\label{rad-density}
\ee
where $R_{\mu \nu}$ is described by (\ref{Ricci-new}).
%

It is of interest to find out whether the MPD equations have any effect on the radiation density, as seen by
a local stationary observer.
To find out, a plot of (\ref{rad-density}) is presented in Figure~\ref{fig:rad-density} for the case of the hyperbolic
tangent mass function.
From comparing Figures~\ref{fig:rad-density:th=0} and~\ref{fig:rad-density:th=pi} for initial spin orientations of
$\hat{\th} = 0$ and $\hat{\th} = \pi$, respectively, it is shown that the MPD spin-gravity coupling induces about a \mbox{3\%}
shift in the measured radiation density with respect to the maximum value determined by geodesic motion.
It happens that only the component of spin projected along the $z$-direction is sensitive to the difference
in measured radiation density, and the results shown here are consistent with the kinematic effects of the spin orientation
presented earlier.
Because of the velocity dependence inherent in (\ref{rad-density}), the local radiation density is analogous to a velocity-dependent
drag effect with the dimensions of pressure.
However, it is also apparent that with a maximum magnitude of only $q~\approx~6 \times 10^{-10} \, M_0^{-2}$
that rapidly decays to zero, the radiation density of the Vaidya background appears to have little impact on the
spinning particle's orbital motion.

\subsection{Gravitational waveforms and radiation}

\begin{figure*}
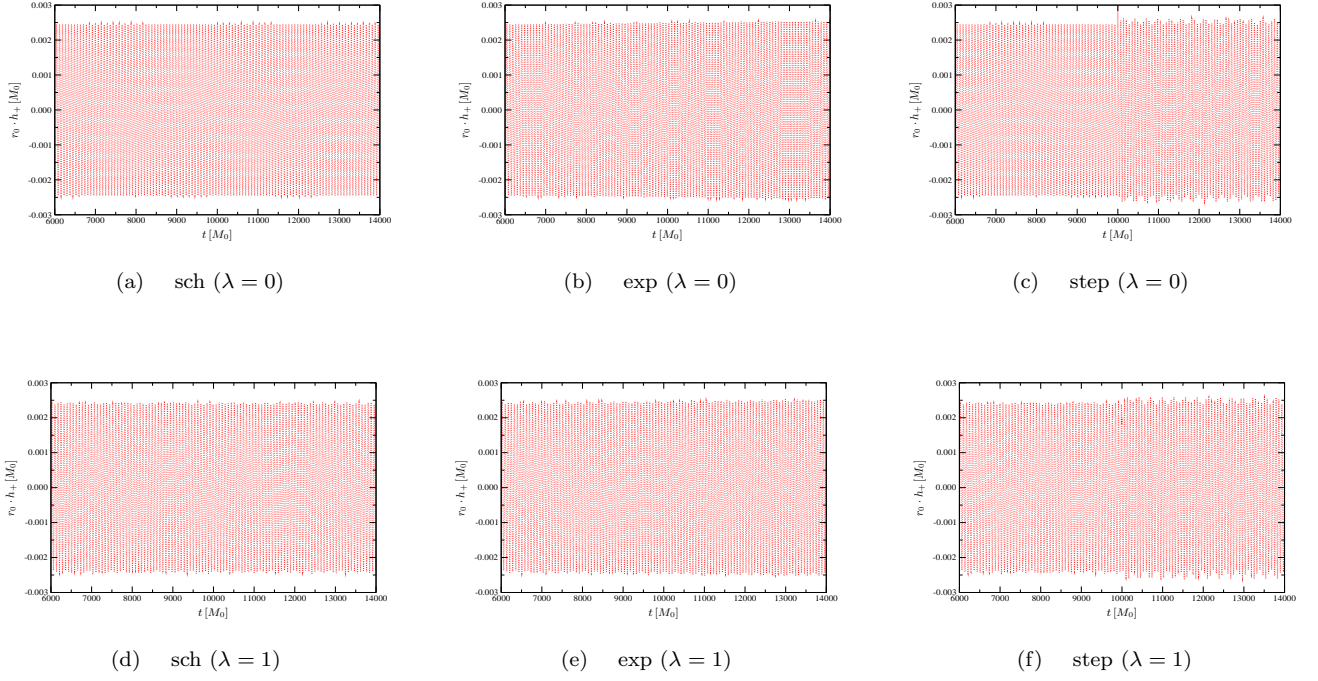

\psfrag{t [M]}[tc][][2.5][0]{$t \, [M_0]$}
\psfrag{r.h_plus [M]}[bc][][2.5][0]{$r_0 \cdot h_+ \, [M_0]$}
\vspace{1mm}
\begin{minipage}[t]{0.3 \textwidth}
\centering
\subfigure[\hspace{0.2cm} sch ($\lm = 0$)]{
\label{fig:h-plus-sch:l=0}
\rotatebox{0}{\includegraphics[width = 5.0cm, height = 3.1cm, scale = 1]{h_plus-new_sch_l=0_m=100E-2_sm=500E-2_v=350E-1_theta=025_phi=025}}}
\end{minipage}%
\hspace{0.5cm}
\begin{minipage}[t]{0.3 \textwidth}
\centering
\subfigure[\hspace{0.2cm} exp ($\lm = 0$)]{
\label{fig:h-plus-exp:l=0-e=001}
\rotatebox{0}{\includegraphics[width = 5.0cm, height = 3.1cm, scale = 1]{h_plus-new_exp_l=0_m=100E-2_sm=500E-2_v=350E-1_theta=025_phi=025}}}
\end{minipage}%
\hspace{0.5cm}
\begin{minipage}[t]{0.3 \textwidth}
\centering
\subfigure[\hspace{0.2cm} step ($\lm = 0$)]{
\label{fig:h-plus-stp:l=0-e=001}
\rotatebox{0}{\includegraphics[width = 5.0cm, height = 3.1cm, scale = 1]{h_plus-new_stp_l=0_m=100E-2_sm=500E-2_v=350E-1_theta=025_phi=025}}}
\end{minipage}
\begin{minipage}[t]{0.3 \textwidth}
\vspace{4mm}
\centering
\subfigure[\hspace{0.2cm} sch ($\lm = 1$)]{
\label{fig:h-plus-sch:l=1}
\rotatebox{0}{\includegraphics[width = 5.0cm, height = 3.1cm, scale = 1]{h_plus-new_sch_l=1_m=100E-2_sm=500E-2_v=350E-1_theta=025_phi=025}}}
\end{minipage}%
\hspace{0.5cm}
\begin{minipage}[t]{0.3 \textwidth}
\vspace{4mm}
\centering
\subfigure[\hspace{0.2cm} exp ($\lm = 1$)]{
\label{fig:h-plus-exp:l=1-e=001}
\rotatebox{0}{\includegraphics[width = 5.0cm, height = 3.1cm, scale = 1]{h_plus-new_exp_l=1_m=100E-2_sm=500E-2_v=350E-1_theta=025_phi=025}}}
\end{minipage}
\hspace{0.5cm}
\begin{minipage}[t]{0.3 \textwidth}
\vspace{4mm}
\centering
\subfigure[\hspace{0.2cm} step ($\lm = 1$)]{
\label{fig:h-plus-stp:l=1-e=001}
\rotatebox{0}{\includegraphics[width = 5.0cm, height = 3.1cm, scale = 1]{h_plus-new_stp_l=1_m=100E-2_sm=500E-2_v=350E-1_theta=025_phi=025}}}
\end{minipage}
\caption{\label{fig:h-plus-e=001} Waveforms in the $+$~polarization mode for Schwarzschild, logarithmic, and step mass functions,
where $\varepsilon = 0.01.$  The initial spin orientation for all cases is $\hat{\th} = \hat{\ph} = \pi/4$.
The effects of MPD spin-gravity coupling shows that a distinctive modulation appears in the waveform peaks.
\vspace{2mm}}
\end{figure*}
The next step in this Section is to obtain the gravitational waveforms and generate the radiation output of
the spinning particle while in orbit around the accreting central mass.
For this exercise, the $+$ and $\times$ polarization modes are described by the transverse traceless (TT) gauge,
where the reduced quadrupole moment formula is assumed throughout for all relevant calculations.
Although this assumption is most likely not sufficient for the purposes of comparing with potential observations by LIGO or LISA,
where a more sophisticated analysis in terms of black hole perturbation theory may be required, it is still
possible to gain valuable qualitative information about the waveforms' leading-order properties.

To begin, the reduced quadrupole moment \cite{Suzuki2} for the point particle is
\be
Q_{ij} & = & m \lt(x_i \, x_j - {1 \over 3} \, \delta_{ij} \, r^2 \rt),
\label{Q-ij}
\ee
where $x_i$ is the particle's Cartesian co-ordinate in space relative to the central mass.
Before proceeding further, it should be noted that (\ref{Q-ij}) is defined for co-ordinates
that are isotropic in space, where the tensor indices are raised and lowered with the Minkowski metric.
Because the Vaidya metric as represented by (\ref{Vaidya-new}) is not isotropic in space, there will be
some lack of precision in the gravitational waveform amplitudes due to corrections in the
radial and/or time co-ordinate to generate a more suitably defined metric.
Using as a guideline the co-ordinate transformation \cite{Singh}
\begin{subequations}
\label{r-isotropic}
\be
\bar{r}(r) & = & {r \over 2} \lt[1 - {M \over r} + \lt(1 - {2M \over r}\rt)^{1/2}\rt]
\nl
r(\bar{r}) & = & \lt(1 + {M \over 2\bar{r}}\rt)^2 \bar{r}
\ee
\end{subequations}
to describe the Schwarzschild metric in terms of isotropic spherical co-ordinates $(\bar{r}, \th, \ph)$,
it follows that $\bar{r} \approx 7 \, M_0$ for $r \approx 8 \, M_0$, with an error of approximately 12\%.
Because it appears unlikely that (\ref{Vaidya-new}) can be put into isotropic form in a simple manner,
and given the likelihood that LIGO's or LISA's signal-to-noise ratio will not notice the difference, the
waveforms presented below offer a reasonably suitable qualitative description of the gravitational
radiation generated by the spinning particle.

In the TT gauge and for an observer located at $\lt(r_0, \th_0, \ph_0\rt)$,
the $+$ and $\times$ polarization mode amplitudes for the quadrupole moment formula are
\begin{subequations}
\label{h}
\be
h_+ & = & \lt({\cos^2 \th_0 + 1 \over 4}\rt) \lt[\cos 2\ph_0 \lt(h_{xx} - h_{yy}\rt) \rt.
\nn
& &{} + \lt. 2 \, \sin 2\ph_0 \, h_{xy}\rt]
- {1 \over 4} \, \sin^2 \th_0 \lt(h_{xx} + h_{yy} - h_{zz} \rt)
\nn
& &{} - {1 \over 2} \, \sin 2\th_0 \lt(\cos \ph_0 \, h_{xz} + \sin \ph_0 \, h_{yz}\rt),
\label{h-+}
\nl
h_\times & = & - {1 \over 2} \, \cos \th_0 \lt[\sin 2\ph_0 \lt(h_{xx} - h_{yy}\rt) - 2 \, \cos 2\ph_0 \, h_{xy}\rt]
\nn
& &{}+ \sin \th_0 \lt(\sin \ph_0 \, h_{xz} + \cos \ph_0 \, h_{yz}\rt),
\label{h-x}
\ee
\end{subequations}
where
\be
h_{ij} & = & {2 \over r_0} \, {d^{\, 2} Q_{ij} \over dt^2}.
\label{h-ij}
\ee
\begin{figure*}
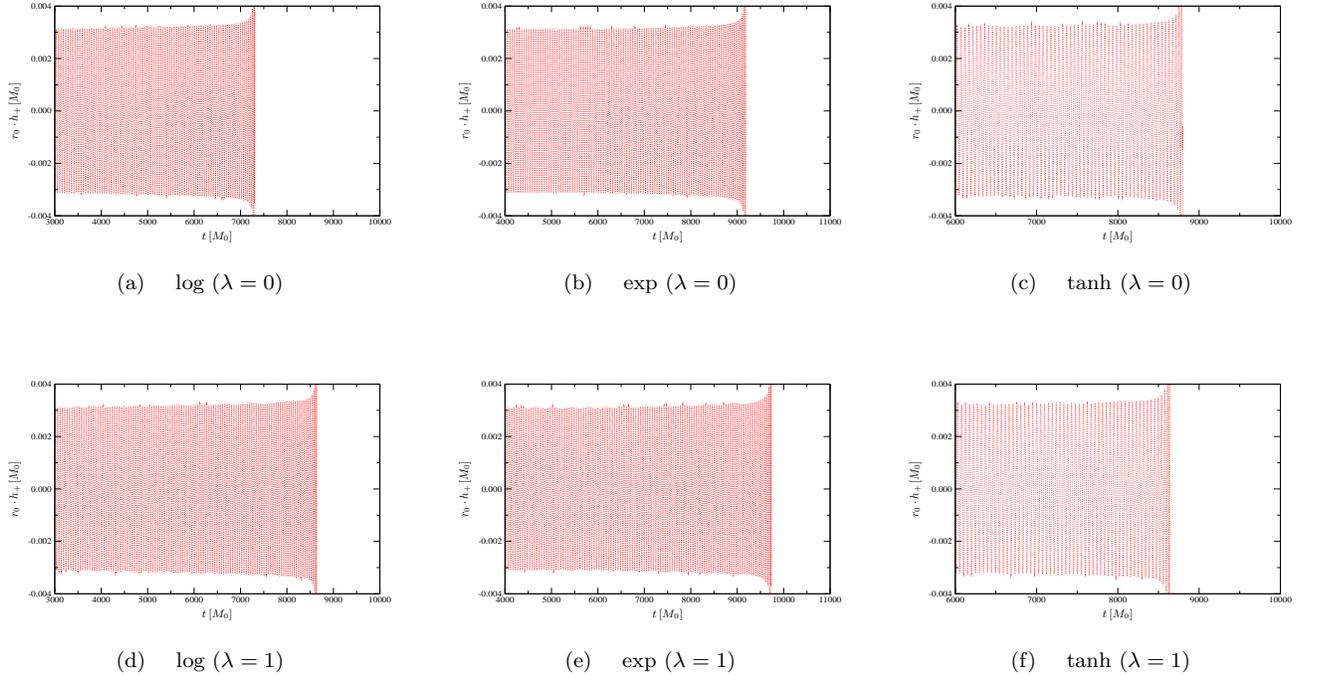

\psfrag{t [M]}[tc][][2.5][0]{$t \, [M_0]$}
\psfrag{r.h_plus [M]}[bc][][2.5][0]{$r_0 \cdot h_+ \, [M_0]$}
\vspace{1mm}
\begin{minipage}[t]{0.3 \textwidth}
\centering
\subfigure[\hspace{0.2cm} log ($\lm = 0$)]{
\label{fig:h-plus-log:l=0-e=005}
\rotatebox{0}{\includegraphics[width = 5.0cm, height = 3.1cm, scale = 1]{h_plus-new_log_l=0_m=100E-2_sm=500E-3_v=395E-1_theta=025_phi=025}}}
\end{minipage}%
\hspace{0.5cm}
\begin{minipage}[t]{0.3 \textwidth}
\centering
\subfigure[\hspace{0.2cm} exp ($\lm = 0$)]{
\label{fig:h-plus-exp:l=0-e=005}
\rotatebox{0}{\includegraphics[width = 5.0cm, height = 3.1cm, scale = 1]{h_plus-new_exp_l=0_m=100E-2_sm=500E-3_v=395E-1_theta=025_phi=025}}}
\end{minipage}%
\hspace{0.5cm}
\begin{minipage}[t]{0.3 \textwidth}
\centering
\subfigure[\hspace{0.2cm} tanh ($\lm = 0$)]{
\label{fig:h-plus-tan:l=0-e=005}
\rotatebox{0}{\includegraphics[width = 5.0cm, height = 3.1cm, scale = 1]{h_plus-new_tan_l=0_m=100E-2_sm=500E-3_v=395E-1_theta=025_phi=025}}}
\end{minipage}
\begin{minipage}[t]{0.3 \textwidth}
\vspace{4mm}
\centering
\subfigure[\hspace{0.2cm} log ($\lm = 1$)]{
\label{fig:h-plus-log:l=1-e=005}
\rotatebox{0}{\includegraphics[width = 5.0cm, height = 3.1cm, scale = 1]{h_plus-new_log_l=1_m=100E-2_sm=500E-3_v=395E-1_theta=025_phi=025}}}
\end{minipage}%
\hspace{0.5cm}
\begin{minipage}[t]{0.3 \textwidth}
\vspace{4mm}
\centering
\subfigure[\hspace{0.2cm} exp ($\lm = 1$)]{
\label{fig:h-plus-exp:l=1-e=005}
\rotatebox{0}{\includegraphics[width = 5.0cm, height = 3.1cm, scale = 1]{h_plus-new_exp_l=1_m=100E-2_sm=500E-3_v=395E-1_theta=025_phi=025}}}
\end{minipage}%
\hspace{0.5cm}
\begin{minipage}[t]{0.3 \textwidth}
\vspace{4mm}
\centering
\subfigure[\hspace{0.2cm} tanh ($\lm = 1$)]{
\label{fig:h-plus-tan:l=1-e=005}
\rotatebox{0}{\includegraphics[width = 5.0cm, height = 3.1cm, scale = 1]{h_plus-new_tan_l=1_m=100E-2_sm=500E-3_v=395E-1_theta=025_phi=025}}}
\end{minipage}
\caption{\label{fig:h-plus-e=005} Waveforms in the $+$~polarization mode for logarithmic and exponential mass functions,
where $\varepsilon = 0.01$, $\hat{\th} = \hat{\ph} = \pi/4$, $v = 0.395 \, c$, and ${\cal S}/m = 5 \times 10^{-3} \, M_0$.
A well-defined amplitude modulation due to MPD spin-gravity coupling is present in
Figs.\ref{fig:h-plus-log:l=1-e=005}--\ref{fig:h-plus-tan:l=1-e=005} during the final stages of particle inspiral.
\vspace{1mm}}
\end{figure*}
For the purpose of this paper, the observer is located along the orbital plane in the $x$-direction,
where $\th_0 = \pi/2$ and $\ph_0 = 0$.
From this vantage point, any particle whose motion is confined to the orbital plane will only generate $h_+$.
Therefore, any observation of $h_\times$ with significant amplitude is a certain indicator of a force
applied on the particle along the polar direction.

\begin{figure*}
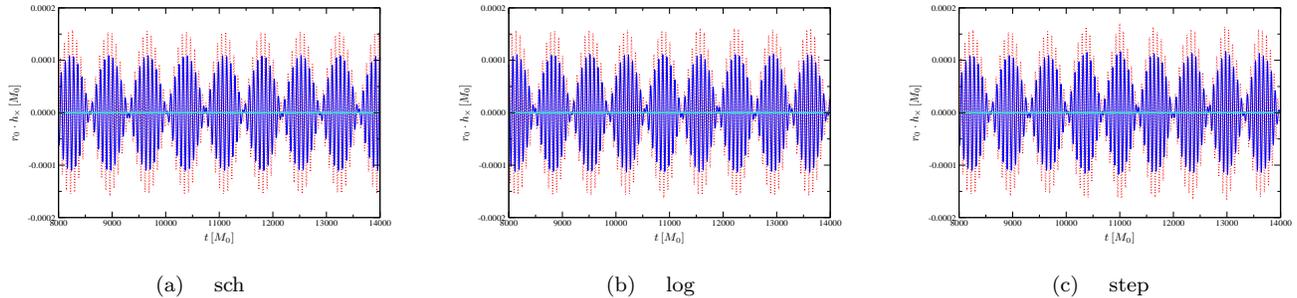

\psfrag{t [M]}[tc][][2.5][0]{$t \, [M_0]$}
\psfrag{r.h_cross [M]}[bc][][2.5][0]{$r_0 \cdot h_\times \, [M_0]$}
\psfrag{t_h = Pi/2, p_h = Pi/2}[cc][][1.98][0]{$\hat{\th} = \pi/2, \, \hat{\ph} = \pi/2$}
\psfrag{t_h = Pi/2, p_h = 0}[cc][][1.98][0]{$\hat{\th} = \pi/2, \, \hat{\ph} = 0$}
\psfrag{t_h = 0}[cc][][1.98][0]{$\hat{\th} = 0$}
\vspace{1mm}
\begin{minipage}[t]{0.3 \textwidth}
\centering
\subfigure[\hspace{0.2cm} sch]{
\label{fig:h-cross-sch}
\rotatebox{0}{\includegraphics[width = 5.0cm, height = 3.1cm, scale = 1]{h_cross-new_sch_l=1_m=100E-2_sm=500E-2_v=350E-1_angles}}}
\end{minipage}%
\hspace{0.5cm}
\begin{minipage}[t]{0.3 \textwidth}
\centering
\subfigure[\hspace{0.2cm} log]{
\label{fig:h-cross-log:e=001}
\rotatebox{0}{\includegraphics[width = 5.0cm, height = 3.1cm, scale = 1]{h_cross-new_log_l=1_m=100E-2_sm=500E-2_v=350E-1_angles}}}
\end{minipage}%
\hspace{0.5cm}
\begin{minipage}[t]{0.3 \textwidth}
\centering
\subfigure[\hspace{0.2cm} step]{
\label{fig:h-cross-stp:e=001}
\rotatebox{0}{\includegraphics[width = 5.0cm, height = 3.1cm, scale = 1]{h_cross-new_stp_l=1_m=100E-2_sm=500E-2_v=350E-1_angles}}}
\end{minipage}
\caption{\label{fig:h-cross-e=001} Waveforms in the $\times$~polarization mode for Schwarzschild, logarithmic, and step mass functions,
where $\varepsilon = 0.01.$  The dotted line representing the largest amplitude corresponds to $\hat{\th} = \pi/2$ and $\hat{\ph} = 0$,
the thin solid line for the second largest amplitude refers to $\hat{\th} = \hat{\ph} = \pi/2$, and the thick solid horizontal line
represents $\hat{\th} = 0$.
\vspace{1mm}}
\end{figure*}
\begin{figure*}
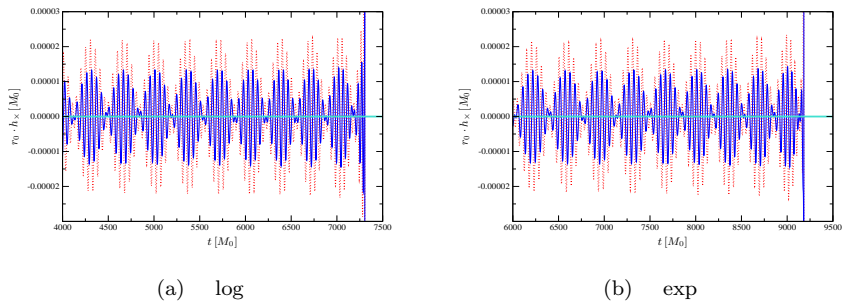

\psfrag{t [M]}[tc][][2.5][0]{$t \, [M_0]$}
\psfrag{r.h_cross [M]}[bc][][2.5][0]{$r_0 \cdot h_\times \, [M_0]$}
\psfrag{t_h = Pi/2, p_h = Pi/2}[cc][][1.98][0]{$\hat{\th} = \pi/2, \, \hat{\ph} = \pi/2$}
\psfrag{t_h = Pi/2, p_h = 0}[cc][][1.98][0]{$\hat{\th} = \pi/2, \, \hat{\ph} = 0$}
\psfrag{t_h = 0}[cc][][1.98][0]{$\hat{\th} = 0$}
\vspace{1mm}
\begin{minipage}[t]{0.3 \textwidth}
\centering
\subfigure[\hspace{0.2cm} log]{
\label{fig:h-cross-log:e=005}
\rotatebox{0}{\includegraphics[width = 5.0cm, height = 3.1cm, scale = 1]{h_cross-new_log_l=1_m=100E-2_sm=500E-3_v=395E-1_angles}}}
\end{minipage}%
\hspace{0.5cm}
\begin{minipage}[t]{0.3 \textwidth}
\centering
\subfigure[\hspace{0.2cm} exp]{
\label{fig:h-cross-exp:e=005}
\rotatebox{0}{\includegraphics[width = 5.0cm, height = 3.1cm, scale = 1]{h_cross-new_exp_l=1_m=100E-2_sm=500E-3_v=395E-1_angles}}}
\end{minipage}%
\hspace{0.5cm}
\begin{minipage}[t]{0.3 \textwidth}
\centering
\vspace{-2.9cm}
\caption{\label{fig:h-cross-e=005} Waveforms in the $\times$~polarization mode for logarithmic and exponential mass functions,
where $\varepsilon~=~0.01$, $\hat{\th} = \hat{\ph} = \pi/4$, $v = 0.395 \, c$, and ${\cal S}/m = 5 \times 10^{-3} \, M_0$.
\vspace{1mm}}
\end{minipage}
\end{figure*}
To compare the contributions of MPD spin-gravity coupling to the $+$ polarization mode,
a set of plots for $h_+$ for the Schwarzschild, exponential, and step mass functions are listed in Figure~\ref{fig:h-plus-e=001},
where $\hat{\th} = \hat{\ph} = \pi/4$ and $\varepsilon = 0.01$.
A review of Figures~\ref{fig:h-plus-sch:l=0} and~\ref{fig:h-plus-sch:l=1} for the Schwarzschild mass shows that
$h_+$ reveals a slightly jagged peak structure in the waveforms when the force and torque terms in the MPD equations
of motion are incorporated, though there is no obvious increase in the waveform's amplitude or frequency of oscillation.

When the effects of mass accretion are included, the particle's orbital response does get reflected in the
$+$ mode, albeit with a very slight increase in the $h_+$ amplitude over time.
Comparison of Figures~\ref{fig:h-plus-exp:l=0-e=001} and~\ref{fig:h-plus-exp:l=1-e=001} with their respective
Schwarzschild counterparts shows how $h_+$ responds to mass accretion.
This is evident in the step mass function examples of Figures~\ref{fig:h-plus-stp:l=0-e=001} and~\ref{fig:h-plus-stp:l=1-e=001},
where it is obvious that a sudden change of mass had occurred at $t = 10^4 \, M_0$.
It is particularly interesting to note for the step function examples that the effect of mass accretion alone is capable of producing a jagged
peak structure in $h_+$ due to purely geodesic motion $(\lm = 0)$ that is virtually indistinguishable from the corresponding
plot when MPD spin-gravity coupling is incorporated.
A short spike does appear in the amplitude of Figure~\ref{fig:h-plus-stp:l=1-e=001} when the mass accretion rate is almost infinite.
Although it is possibly a numerical artifact only, the amplitude spike may be a legitimate signal due to MPD spin-gravity coupling,
since the step mass function as defined from (\ref{tanh}) is formally smooth.

\begin{figure*}
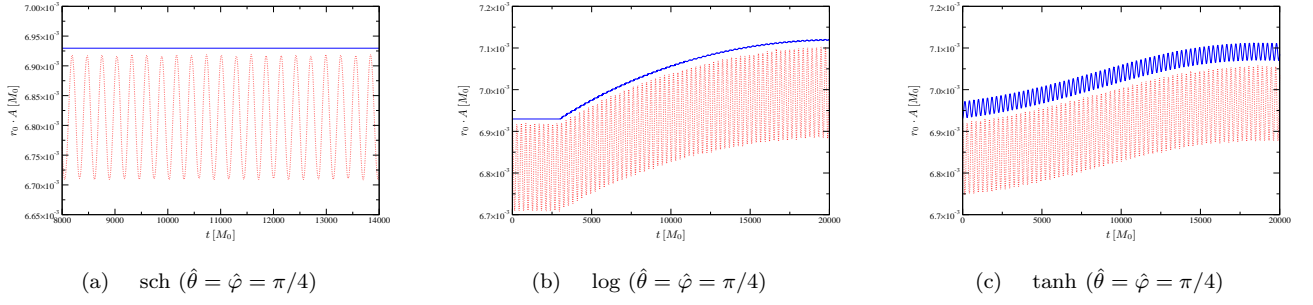

\psfrag{t [M]}[tc][][2.5][0]{$t \, [M_0]$}
\psfrag{r.A [M]}[bc][][2.5][0]{$r_0 \cdot A \, [M_0]$}
\psfrag{l = 1}[cc][][1.5][0]{$\lm = 1$}
\psfrag{l = 0}[cc][][1.5][0]{$\lm = 0$}
\vspace{1mm}
\begin{minipage}[t]{0.3 \textwidth}
\centering
\subfigure[\hspace{0.2cm} sch ($\hat{\th} = \hat{\ph} = \pi/4$)]{
\label{fig:amplitude-sch}
\rotatebox{0}{\includegraphics[width = 5.0cm, height = 3.1cm, scale = 1]{amplitude-new_sch_m=100E-2_sm=500E-2_v=350E-1_theta=025_phi=025}}}
\end{minipage}%
\hspace{0.5cm}
\begin{minipage}[t]{0.3 \textwidth}
\centering
\subfigure[\hspace{0.2cm} log ($\hat{\th} = \hat{\ph} = \pi/4$)]{
\label{fig:amplitude-log:e=001}
\rotatebox{0}{\includegraphics[width = 5.0cm, height = 3.1cm, scale = 1]{amplitude-new_log_m=100E-2_sm=500E-2_v=350E-1_theta=025_phi=025}}}
\end{minipage}%
\hspace{0.5cm}
\begin{minipage}[t]{0.3 \textwidth}
\centering
\subfigure[\hspace{0.2cm} tanh ($\hat{\th} = \hat{\ph} = \pi/4$)]{
\label{fig:amplitude-tan:e=001}
\rotatebox{0}{\includegraphics[width = 5.0cm, height = 3.1cm, scale = 1]{amplitude-new_tan_m=100E-2_sm=500E-2_v=350E-1_theta=025_phi=025}}}
\end{minipage}
\caption{\label{fig:amplitude-e=001} Gravitational wave amplitude for Schwarzschild and other mass functions, averaged
over all directions.
As shown by Figs. ~\ref{fig:amplitude-log:e=001} and \ref{fig:amplitude-tan:e=001}, the amplitude increases
in direct response to the mass function's rate of growth.
The MPD spin-gravity coupling slightly reduces the amplitude compared to the respective value for geodesic motion for each case.
\vspace{1mm}}
\end{figure*}
A more interesting set of plots is found in Figure~\ref{fig:h-plus-e=005}, which captures the $h_+$ waveform of a particle in
the final stages of its inspiralling orbit.
For both the logarithmic and exponential mass functions considered, where $v = 0.395 \, c$, ${\cal S}/m = 5 \times 10^{-3} \, M_0$,
and $\hat{\th} = \hat{\ph} = \pi/4$,
the familiar-looking ``chirp'' signal appears during the particle's final moments before its demise.
Comparison of Figures~\ref{fig:h-plus-log:l=1-e=005}--~\ref{fig:h-plus-tan:l=1-e=005} with their respective plots for $\lambda = 0$
suggests the presence of a modulation in the amplitude before final inspiral due to MPD spin-gravity coupling.
However, the degree of modulation is very slight for these examples, and it is unlikely to be able to identify the
presence of particle spin in $h_+$, given the precision required for LIGO or LISA to identify a candidate signal.

In contrast, of potentially much greater relevance for this paper is the effect of MPD spin-gravity coupling on the
$h_\times$ waveform and the correlation of amplitude with its spin orientation.
Because the particle oscillates about $\th = \pi/2$, some $\times$ polarization radiation must be emitted,
though it is not obvious about the amplitude strength when compared to $h_+$, as seen by an observer on the orbital plane.
Figure~\ref{fig:h-cross-e=001} lists the $\times$ mode waveforms for the Schwarzschild, logarithmic, and step mass functions,
where the dotted line corresponds to initial spin orientation $\hat{\th} = \pi/2$ and
$\hat{\ph} =  0$, the thin solid line describes $\hat{\th} = \hat{\ph} = \pi/2$,
and the thick solid horizontal line represents $\hat{\th} = 0$.
For the case of an inspiralling particle, Figure~\ref{fig:h-cross-e=005} list the plots of $h_\times$ for the logarithmic and
exponential functions, where $v = 0.395 \, c$, and ${\cal S}/m = 5 \times 10^{-3} \, M_0$.
It is interesting to note that the maximum amplitude of $h_\times$ is about \mbox{5\%} of the corresponding amplitude for
$h_+$, as described by Figure~\ref{fig:h-plus-e=001}.
While the projection of spin orientation along the radial direction of motion ($\hat{\th} = \pi/2$ and $\hat{\ph} = 0$)
contributes the largest amplitude, the contribution along the tangential direction ($\hat{\th} = \hat{\ph} = \pi/2$) is
comparably large, as it generates about \mbox{55\%} of the maximum amplitude.
For spin orientation along the positive normal direction ($\hat{\th} = 0$), there is no $\times$ mode radiation emitted.
These results suggests the interpretation that, over time, the particle is trying to align its spin in the positive
$z$-direction as a point of stable equilibrium, reinforcing the same claim made earlier based on other data.

\begin{figure*}
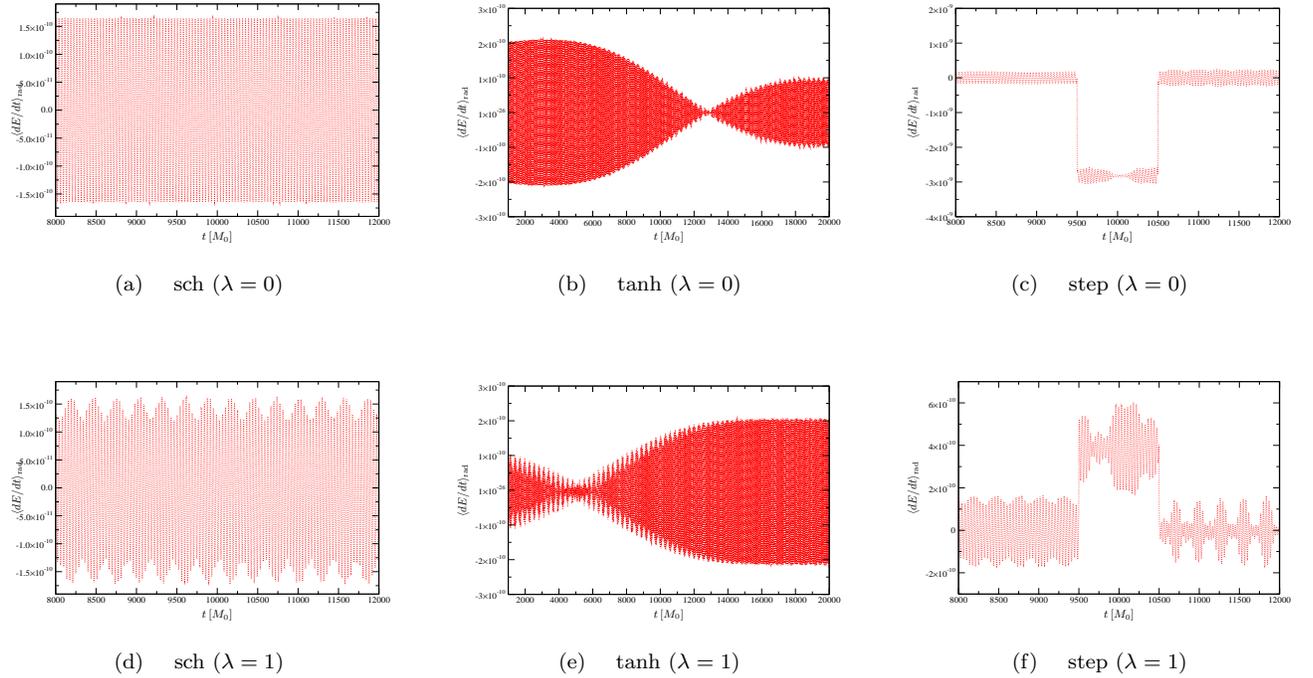

\psfrag{t [M]}[tc][][2.5][0]{$t \, [M_0]$}
\psfrag{<dE/dt>}[bc][][2.5][0]{$\langle dE/dt \rangle_{\rm rad}$}
\vspace{1mm}
\begin{minipage}[t]{0.3 \textwidth}
\centering
\subfigure[\hspace{0.2cm} sch ($\lm = 0$)]{
\label{fig:energy-sch:l=0}
\rotatebox{0}{\includegraphics[width = 5.0cm, height = 3.1cm, scale = 1]{energy-new_sch_l=0_m=100E-2_sm=500E-2_v=350E-1_theta=025_phi=025}}}
\end{minipage}%
\hspace{0.5cm}
\begin{minipage}[t]{0.3 \textwidth}
\centering
\subfigure[\hspace{0.2cm} tanh ($\lm = 0$)]{
\label{fig:energy-tan:l=0-e=001}
\rotatebox{0}{\includegraphics[width = 5.0cm, height = 3.1cm, scale = 1]{energy-new_tan_l=0_m=100E-2_sm=500E-2_v=350E-1_theta=025_phi=025}}}
\end{minipage}%
\hspace{0.5cm}
\begin{minipage}[t]{0.3 \textwidth}
\centering
\subfigure[\hspace{0.2cm} step ($\lm = 0$)]{
\label{fig:energy-stp:l=0-e=001}
\rotatebox{0}{\includegraphics[width = 5.0cm, height = 3.1cm, scale = 1]{energy-new_stp_l=0_m=100E-2_sm=500E-2_v=350E-1_theta=025_phi=025}}}
\end{minipage}
\begin{minipage}[t]{0.3 \textwidth}
\vspace{4mm}
\centering
\subfigure[\hspace{0.2cm} sch ($\lm = 1$)]{
\label{fig:energy-sch:l=1}
\rotatebox{0}{\includegraphics[width = 5.0cm, height = 3.1cm, scale = 1]{energy-new_sch_l=1_m=100E-2_sm=500E-2_v=350E-1_theta=025_phi=025}}}
\end{minipage}%
\hspace{0.5cm}
\begin{minipage}[t]{0.3 \textwidth}
\vspace{4mm}
\centering
\subfigure[\hspace{0.2cm} tanh ($\lm = 1$)]{
\label{fig:energy-tan:l=1-e=001}
\rotatebox{0}{\includegraphics[width = 5.0cm, height = 3.1cm, scale = 1]{energy-new_tan_l=1_m=100E-2_sm=500E-2_v=350E-1_theta=025_phi=025}}}
\end{minipage}%
\hspace{0.5cm}
\begin{minipage}[t]{0.3 \textwidth}
\vspace{4mm}
\centering
\subfigure[\hspace{0.2cm} step ($\lm = 1$)]{
\label{fig:energy-stp:l=1-e=001}
\rotatebox{0}{\includegraphics[width = 5.0cm, height = 3.1cm, scale = 1]{energy-new_stp_l=1_m=100E-2_sm=500E-2_v=350E-1_theta=025_phi=025}}}
\end{minipage}
\caption{\label{fig:energy-e=001} Energy emission rate for Schwarzschild, hyperbolic tangent, and step mass functions,
where $\varepsilon = 0.01$ and $\hat{\th} = \hat{\ph} = \pi/4$.
The MPD spin-gravity coupling has a modulating effect on the emission rate, though for most mass functions considered the
net energy loss is effectively zero.
The only exception is for the step mass function, where there is a non-zero emission at the moment when the mass suddenly increases.
From comparing Fig.~\ref{fig:energy-stp:l=1-e=001} to Fig.~\ref{fig:energy-stp:l=0-e=001}, the MPD spin-gravity coupling
slightly increases the energy emission rate for this choice of spin orientation, though the effect is still negligibly small.
\vspace{1mm}}
\end{figure*}
A view of Figure~\ref{fig:h-cross-sch} describing $h_\times$ for the Schwarzschild mass shows a well-defined and symmetrical
modulation of the amplitude due to the MPD spin gravity coupling.
The effect of mass accretion, as described by Figures~\ref{fig:h-cross-log:e=001} and~\ref{fig:h-cross-stp:e=001}, causes
a slight distortion in the amplitude, which becomes more noticeable with a faster rate of mass accretion.
More significantly, however, the wavetrain gets compressed over the time scale where the mass accretion is most pronounced,
with a more narrow modulation of $h_+$.
This is especially clear from the step function plots in Figure~\ref{fig:h-cross-stp:e=001} where the particle is able to fit
in roughly an extra six or more orbital cycles over the same time scale as compared to Figure~\ref{fig:h-cross-sch}.
This phase shift is also present for the logarithmic function in Figure~\ref{fig:h-cross-log:e=001},
albeit a slightly more modest effect in this case.

The next step is to examine the impact of mass accretion on the predicted energy and orbital angular momentum emission rates for
radiation generated by the spinning point particle.
One area where the impact becomes immediately apparent occurs when plotting the overall amplitude $A$ of the space-averaged
gravitational wave \cite{Suzuki2}, where from (\ref{h-ij}),
\be
A & \equiv & \lt(h_{ij}  \, h^{ij}\rt)^{1/2}
\nn
& = & {2 \over r_0} \lt({d^{\, 2} Q_{ij} \over dt^2} \, {d^{\, 2} Q^{ij} \over dt^2}\rt)^{1/2}.
\label{A}
\ee
Plots of (\ref{A}) scaled with $r_0$ are listed in Figure~\ref{fig:amplitude-e=001}, which describe the
gravitational wave amplitude for each of the mass functions.
From Figure~\ref{fig:amplitude-sch}, the MPD spin gravity coupling has the effect of slightly reducing the overall
amplitude compared to that from purely geodesic motion, while generating a relatively large variation.
When mass accretion is incorporated, it is clear from Figures~\ref{fig:amplitude-log:e=001} and \ref{fig:amplitude-tan:e=001}
that the amplitude responds almost immediately and in proportion to the mass accretion rate.
It is interesting to note that, from Figure~\ref{fig:amplitude-tan:e=001}, the amplitude variation for the $\lm = 0$ plot
is roughly a quarter as for the $\lm = 1$ plot, and the reason for this anomalous behaviour is not entirely clear,
especially when compared to Figure~\ref{fig:amplitude-log:e=001}.
One possible explanation is that the spin-gravity interaction due to parallel transport gradually grows
over time for a smoothly evolving mass function, while this opportunity is not available for the other mass functions considered.
Therefore, it may be a numerical byproduct only.

\begin{figure*}
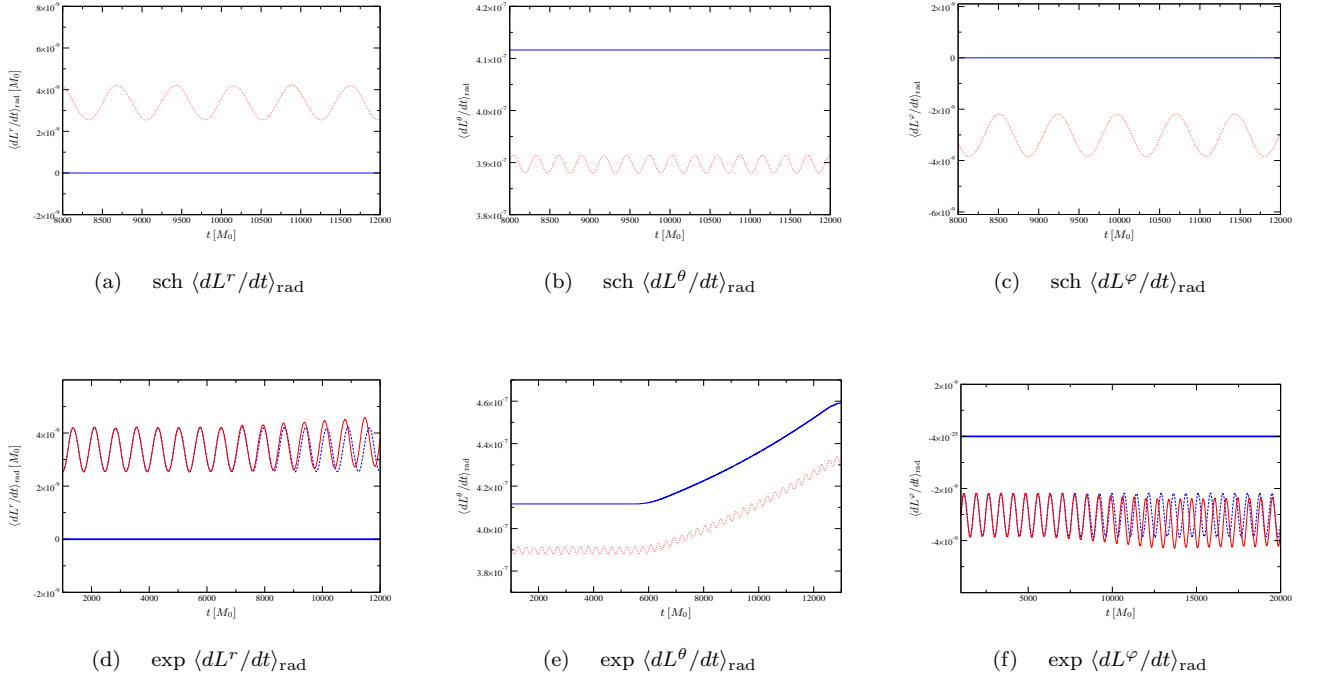

\psfrag{t [M]}[tc][][2.5][0]{$t \, [M_0]$}
\psfrag{<dL^r/dt> [M]}[bc][][2.5][0]{$\langle dL^r/dt \rangle_{\rm rad} \, [M_0]$}
\psfrag{<dL^theta/dt>}[bc][][2.5][0]{$\langle dL^\th/dt \rangle_{\rm rad}$}
\psfrag{<dL^phi/dt>}[bc][][2.5][0]{$\langle dL^\ph/dt \rangle_{\rm rad}$}
\psfrag{l = 1}[cc][][1.5][0]{$\lm = 1$}
\psfrag{l = 0}[cc][][1.5][0]{$\lm = 0$}
\vspace{1mm}
\begin{minipage}[t]{0.3 \textwidth}
\centering
\subfigure[\hspace{0.2cm} sch $\langle dL^r/dt \rangle_{\rm rad}$]{
\label{fig:ang-r-sch}
\rotatebox{0}{\includegraphics[width = 5.0cm, height = 3.1cm, scale = 1]{ang_r-new_sch_m=100E-2_sm=500E-2_v=350E-1_theta=025_phi=025}}}
\end{minipage}%
\hspace{0.5cm}
\begin{minipage}[t]{0.3 \textwidth}
\centering
\subfigure[\hspace{0.2cm} sch $\langle dL^\th/dt \rangle_{\rm rad}$]{
\label{fig:ang-t-sch}
\rotatebox{0}{\includegraphics[width = 5.0cm, height = 3.1cm, scale = 1]{ang_t-new_sch_m=100E-2_sm=500E-2_v=350E-1_theta=025_phi=025}}}
\end{minipage}%
\hspace{0.5cm}
\begin{minipage}[t]{0.3 \textwidth}
\centering
\subfigure[\hspace{0.2cm} sch $\langle dL^\ph/dt \rangle_{\rm rad}$]{
\label{fig:ang-p-sch}
\rotatebox{0}{\includegraphics[width = 5.0cm, height = 3.1cm, scale = 1]{ang_p-new_sch_m=100E-2_sm=500E-2_v=350E-1_theta=025_phi=025}}}
\end{minipage}
\begin{minipage}[t]{0.3 \textwidth}
\vspace{4mm}
\centering
\subfigure[\hspace{0.2cm} exp $\langle dL^r/dt \rangle_{\rm rad}$]{
\label{fig:ang-r-exp:e=001}
\rotatebox{0}{\includegraphics[width = 5.0cm, height = 3.1cm, scale = 1]{ang_r-new_exp_m=100E-2_sm=500E-2_v=350E-1_theta=025_phi=025}}}
\end{minipage}%
\hspace{0.5cm}
\begin{minipage}[t]{0.3 \textwidth}
\vspace{4mm}
\centering
\subfigure[\hspace{0.2cm} exp $\langle dL^\th/dt \rangle_{\rm rad}$]{
\label{fig:ang-t-exp:e=001}
\rotatebox{0}{\includegraphics[width = 5.0cm, height = 3.1cm, scale = 1]{ang_t-new_exp_m=100E-2_sm=500E-2_v=350E-1_theta=025_phi=025}}}
\end{minipage}%
\hspace{0.5cm}
\begin{minipage}[t]{0.3 \textwidth}
\vspace{4mm}
\centering
\subfigure[\hspace{0.2cm} exp $\langle dL^\ph/dt \rangle_{\rm rad}$]{
\label{fig:ang-p-exp:e=001}
\rotatebox{0}{\includegraphics[width = 5.0cm, height = 3.1cm, scale = 1]{ang_p-new_exp_m=100E-2_sm=500E-2_v=350E-1_theta=025_phi=025}}}
\end{minipage}
\caption{\label{fig:ang:e=001} Orbital angular momentum emission rate for Schwarzschild and exponential mass functions,
where $\varepsilon = 0.01$ and $\hat{\th} = \hat{\ph} = \pi/4$.
The dashed oscillating lines in Figs.~\ref{fig:ang-r-exp:e=001} and \ref{fig:ang-p-exp:e=001} correspond to
radiated orbital angular momentum of the particle orbiting a Schwarzschild black hole.
\vspace{1mm}}
\end{figure*}
It is of interest to determine how much energy and orbital angular momentum gets radiated away by the spinning particle in orbit
around an accreting mass.
Based on the quadrupole moment formula, the energy emission rate and the Cartesian component of the
orbital angular momentum emission rate are \cite{Suzuki2}
\be
\lt\langle{dE \over dt}\rt\rangle_{\rm rad}
& = & {1 \over 5} \lt\langle {d^{\, 3} Q_{ij} \over dt^3} \, {d^{\, 3} Q^{ij} \over dt^3} \rt\rangle,
\label{E-emission}
\nl
\lt\langle{dL^i \over dt}\rt\rangle_{\rm rad}
& = & {2 \over 5} \, \epsilon^{ijk} \lt\langle {d^{\, 2} Q_{jm} \over dt^2} \, {d^{\, 3} Q^m{}_k \over dt^3} \rt\rangle,
\label{L-emission}
\ee
where these expressions are time-averaged over several cycles.
The sign conventions adopted for (\ref{E-emission}) and (\ref{L-emission}) are that positive-valued energy and angular momentum
are gained by the system.

\begin{figure*}
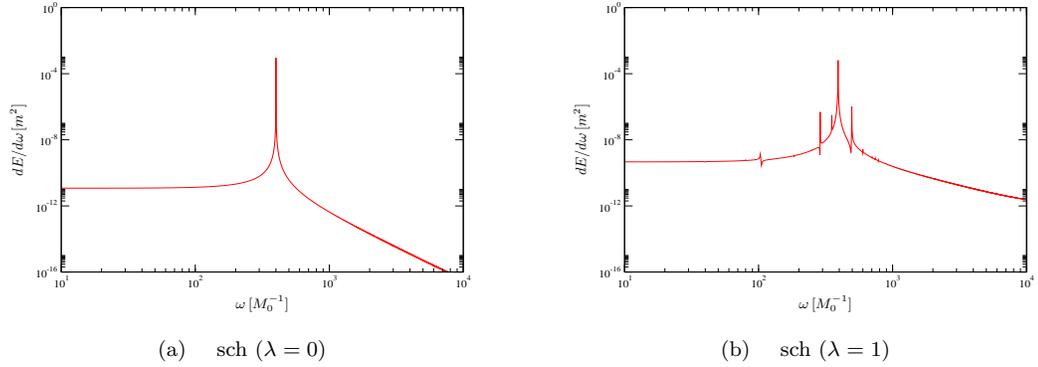

\psfrag{omega [M^(-1)]}[tc][][2.5][0]{$\omega \, [M_0^{-1}]$}
\psfrag{dE/d(omega) [m^2]}[bc][][2.5][0]{$dE/d\omega \, [m^2]$}
\vspace{1mm}
\begin{minipage}[t]{0.3 \textwidth}
\centering
\subfigure[\hspace{0.2cm} sch ($\lm = 0$)]{
\label{fig:power-sch:l=0}
\rotatebox{0}{\includegraphics[width = 6.0cm, height = 4.0cm, scale = 1]{power-new_sch_l=0_m=100E-2_sm=500E-2_v=350E-1_theta=025_phi=025}}}
\end{minipage}%
\hspace{2.0cm}
\begin{minipage}[t]{0.3 \textwidth}
\centering
\subfigure[\hspace{0.2cm} sch ($\lm = 1$)]{
\label{fig:power-sch:l=1}
\rotatebox{0}{\includegraphics[width = 6.0cm, height = 4.0cm, scale = 1]{power-new_sch_l=1_m=100E-2_sm=500E-2_v=350E-1_theta=025_phi=025}}}
\end{minipage}%
\caption{\label{fig:power-sch} Power spectra for the Schwarzschild mass function, where $\hat{\th} = \hat{\ph} = \pi/4$.
\vspace{1mm}}
\end{figure*}
Figure~\ref{fig:energy-e=001} lists a set of energy emission rate plots for the Schwarzschild, hyperbolic tangent, and
step mass functions, with spin orientation $\hat{\th} = \hat{\ph} = \pi/4$.
Comparing Figures~\ref{fig:energy-sch:l=0} and~\ref{fig:energy-sch:l=1} for the Schwarzschild mass shows that the
the MPD spin-gravity coupling modulates the emission rate, but remains essentially zero-valued.
Similarly, Figures~\ref{fig:energy-tan:l=0-e=001} and~\ref{fig:energy-tan:l=1-e=001} corresponding to the hyperbolic tangent
mass function shows that while a gradual mass accretion adds some structure to the plot, the net effect on the emission rate
is negligible.
The exception to this rule occurs when the mass accretion rate is very rapid, such as found in
Figures~\ref{fig:energy-stp:l=0-e=001} and~\ref{fig:energy-stp:l=1-e=001} for the step mass function, where
a sudden burst of radiation is emitted from the particle at the moment where the mass increase rate is highest
before settling back to zero emission.
The effect of MPD spin-gravity coupling for this instance is to reduce the emission rate to a practically negligible
magnitude.

The presence of particle spin adds an interesting dimension to the decay of orbital angular momentum from gravitational
radiation.
This is evident from examining Figure~\ref{fig:ang:e=001}, which details the time-averaged emission rate along all three
spatial directions due to the Schwarzschild and exponential mass functions, assuming $\varepsilon = 0.01$ and
$\hat{\th} = \hat{\ph} = \pi/4$ for the initial spin orientation.
Not surprisingly, the loss of orbital angular momentum is overwhelmingly directed along the $\th$-direction, which is normal
to the $\th = \pi/2$ orbital plane, where the emission is two orders of magnitude higher than for the contributions
along the radial and azimuthal directions.
(It is important to remember that, for a particle in counterclockwise orbit along the $\th = \pi/2$ plane, the orbital angular
momentum is directed along the {\em negative} $\th$-direction.)

\begin{figure*}
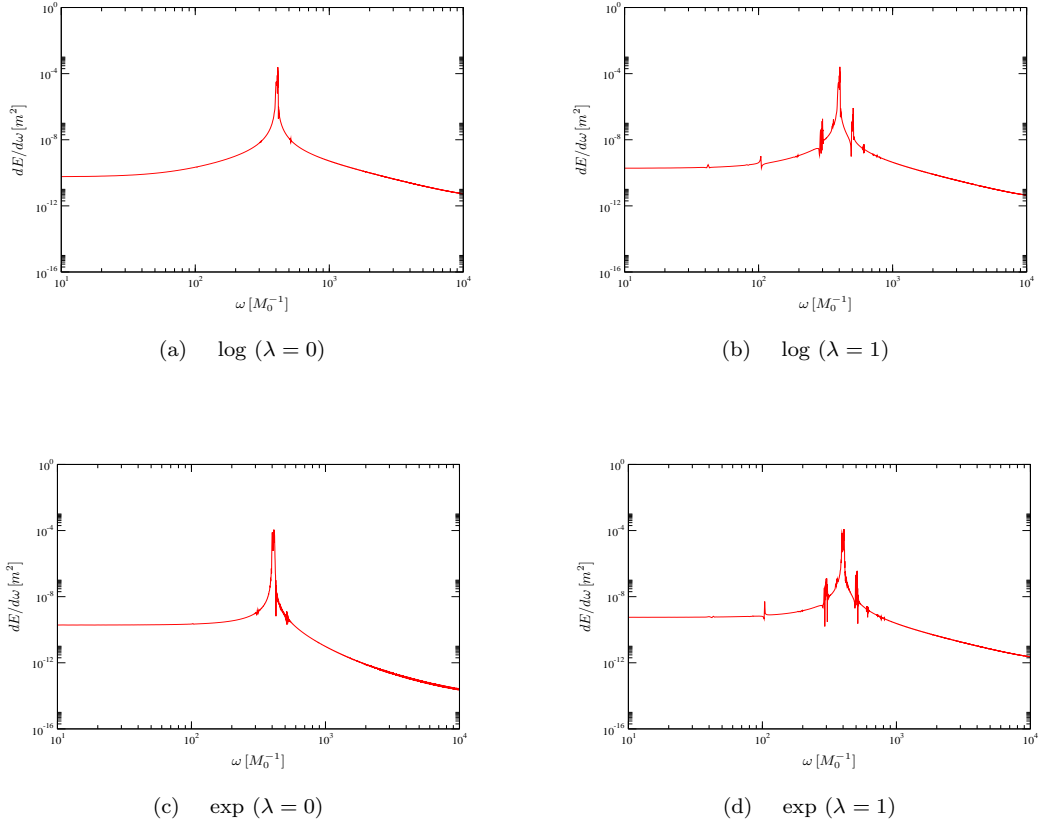

\psfrag{omega [M^(-1)]}[tc][][2.5][0]{$\omega \, [M_0^{-1}]$}
\psfrag{dE/d(omega) [m^2]}[bc][][2.5][0]{$dE/d\omega \, [m^2]$}
\vspace{1mm}
\begin{minipage}[t]{0.3 \textwidth}
\centering
\subfigure[\hspace{0.2cm} log ($\lm = 0$)]{
\label{fig:power-log:l=0-e=001}
\rotatebox{0}{\includegraphics[width = 6.0cm, height = 4.0cm, scale = 1]{power-new_log_l=0_m=100E-2_sm=500E-2_v=350E-1_theta=025_phi=025}}}
\end{minipage}%
\hspace{2.0cm}
\begin{minipage}[t]{0.3 \textwidth}
\centering
\subfigure[\hspace{0.2cm} log ($\lm = 1$)]{
\label{fig:power-log:l=1-e=001}
\rotatebox{0}{\includegraphics[width = 6.0cm, height = 4.0cm, scale = 1]{power-new_log_l=1_m=100E-2_sm=500E-2_v=350E-1_theta=025_phi=025}}}
\end{minipage}
\begin{minipage}[t]{0.3 \textwidth}
\vspace{5.5mm}
\centering
\subfigure[\hspace{0.2cm} exp ($\lm = 0$)]{
\label{fig:power-exp:l=0-e=001}
\rotatebox{0}{\includegraphics[width = 6.0cm, height = 4.0cm, scale = 1]{power-new_exp_l=0_m=100E-2_sm=500E-2_v=350E-1_theta=025_phi=025}}}
\end{minipage}
\hspace{2.0cm}
\begin{minipage}[t]{0.3 \textwidth}
\vspace{5.5mm}
\centering
\subfigure[\hspace{0.2cm} exp ($\lm = 1$)]{
\label{fig:power-exp:l=1-e=001}
\rotatebox{0}{\includegraphics[width = 6.0cm, height = 4.0cm, scale = 1]{power-new_exp_l=1_m=100E-2_sm=500E-2_v=350E-1_theta=025_phi=025}}}
\end{minipage}
\caption{\label{fig:power-log-exp} Power spectra for the logarithmic and exponential mass functions, where $\hat{\th} = \hat{\ph} = \pi/4$.
\vspace{1mm}}
\end{figure*}
\begin{figure*}
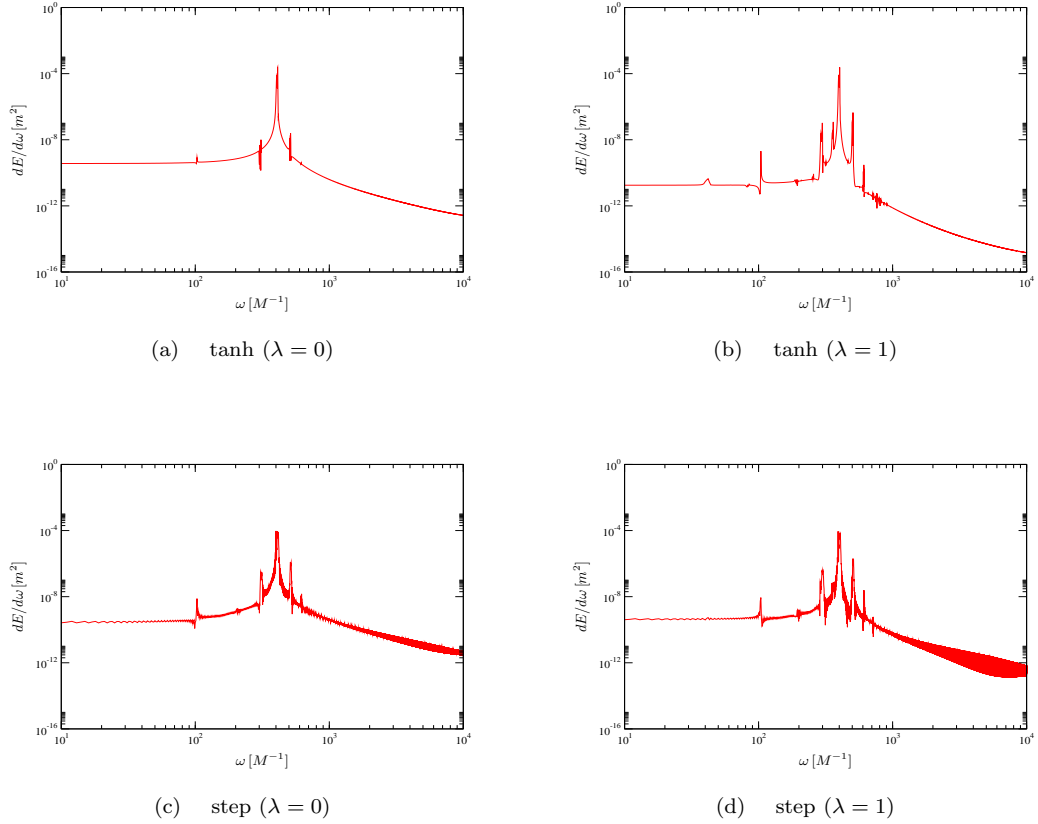

\psfrag{omega [M^(-1)]}[tc][][2.5][0]{$\omega \, [M^{-1}]$}
\psfrag{dE/d(omega) [m^2]}[bc][][2.5][0]{$dE/d\omega \, [m^2]$}
\vspace{1mm}
\begin{minipage}[t]{0.3 \textwidth}
\centering
\subfigure[\hspace{0.2cm} tanh ($\lm = 0$)]{
\label{fig:power-tan:l=0-e=001}
\rotatebox{0}{\includegraphics[width = 6.0cm, height = 4.0cm, scale = 1]{power-new_tan_l=0_m=100E-2_sm=500E-2_v=350E-1_theta=025_phi=025}}}
\end{minipage}%
\hspace{2.0cm}
\begin{minipage}[t]{0.3 \textwidth}
\centering
\subfigure[\hspace{0.2cm} tanh ($\lm = 1$)]{
\label{fig:power-tan:l=1-e=001}
\rotatebox{0}{\includegraphics[width = 6.0cm, height = 4.0cm, scale = 1]{power-new_tan_l=1_m=100E-2_sm=500E-2_v=350E-1_theta=025_phi=025}}}
\end{minipage}
\begin{minipage}[t]{0.3 \textwidth}
\vspace{5.5mm}
\centering
\subfigure[\hspace{0.2cm} step ($\lm = 0$)]{
\label{fig:power-stp:l=0-e=001}
\rotatebox{0}{\includegraphics[width = 6.0cm, height = 4.0cm, scale = 1]{power-new_stp_l=0_m=100E-2_sm=500E-2_v=350E-1_theta=025_phi=025}}}
\end{minipage}%
\hspace{2.0cm}
\begin{minipage}[t]{0.3 \textwidth}
\vspace{5.5mm}
\centering
\subfigure[\hspace{0.2cm} step ($\lm = 1$)]{
\label{fig:power-stp:l=1-e=001}
\rotatebox{0}{\includegraphics[width = 6.0cm, height = 4.0cm, scale = 1]{power-new_stp_l=1_m=100E-2_sm=500E-2_v=350E-1_theta=025_phi=025}}}
\end{minipage}%
\caption{\label{fig:power-tan-stp} Power spectra for hyperbolic tangent and step mass functions,
where $\varepsilon = 0.01$ and $\hat{\th} = \hat{\ph} = \pi/4$.
\vspace{1mm}}
\end{figure*}
Focussing for the moment on Figures~\ref{fig:ang-r-sch}--\ref{fig:ang-p-sch} for the Schwarzschild case,
the first interesting point is that the system gains orbital angular momentum along the radial and azimuthal directions
due to MPD spin-gravity coupling, while there is no corresponding emission for purely geodesic motion.
The $\lm = 1$ plot for Figure~\ref{fig:ang-r-sch} shows that
$\lt|\langle dL^r/dt \rangle_{\rm rad}\rt| \approx 3.5 \times 10^{-9} \, M_0$.
For Figure~\ref{fig:ang-p-sch}, $\lt|r \, \sin \th \, \langle dL^\ph/dt \rangle_{\rm rad}\rt| \approx 2.4 \times 10^{-8} \, M_0$,
roughly an order of magnitude larger.
However, the bulk of the radiation emission is along the $\th$-direction, with magnitude
$\lt|r \, \langle dL^\th/dt \rangle_{\rm rad}\rt| \approx 3.1 \times 10^{-6} \, M_0$ from Figure~\ref{fig:ang-t-sch}.
Over the course of one orbit, the estimated radiation loss is
\be
\Delta L^\th & \approx & \lt\langle {dL^\th \over dt} \rt\rangle_{\rm rad} \Delta t
\nn
& \approx & 5.5 \times 10^{-5} \, M_0,
\label{dL-rad}
\ee
where $\Delta t \simeq 1.4 \times 10^2 \, M_0$ for a single orbit.
By comparison, $L^\th = -4.16 \times 10^{-3} \, M_0$ and $S^\th = -6 \times 10^{-5} \, M_0$ for the spinning particle, so
(\ref{dL-rad}) radiates away a magnitude of angular momentum comparable to that of the particle's spin angular momentum.

\begin{figure*}
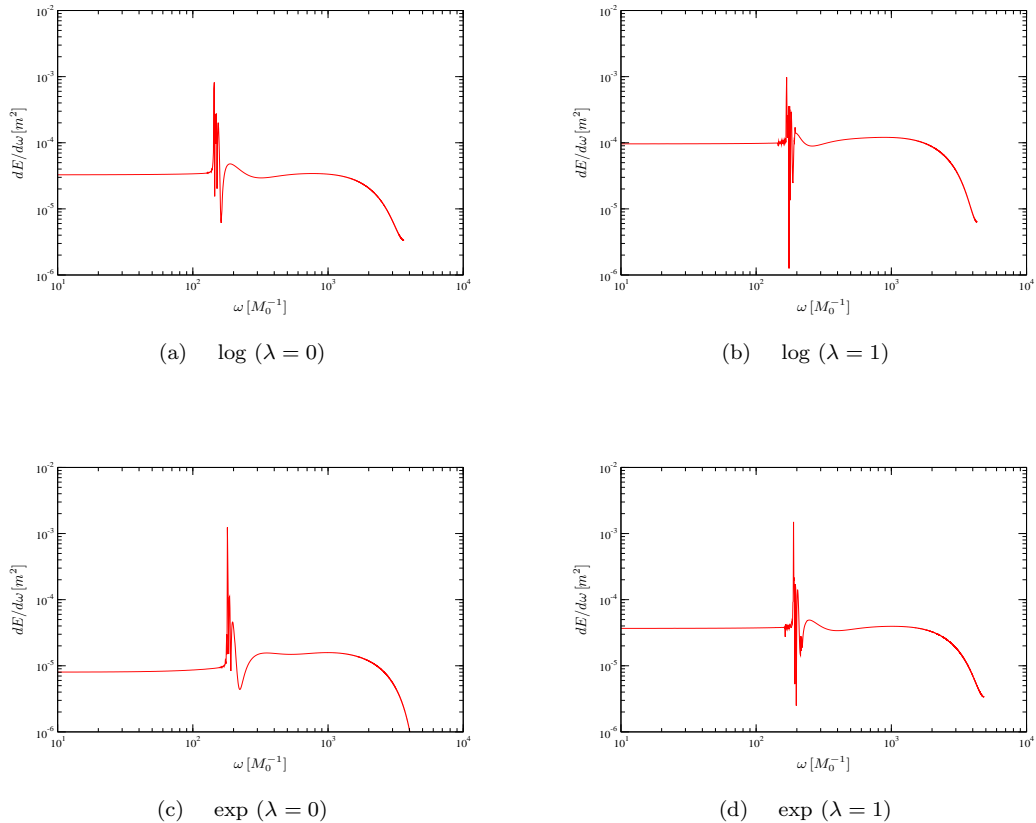

\psfrag{omega [M^(-1)]}[tc][][2.5][0]{$\omega \, [M_0^{-1}]$}
\psfrag{dE/d(omega) [m^2]}[bc][][2.5][0]{$dE/d\omega \, [m^2]$}
\vspace{1mm}
\begin{minipage}[t]{0.3 \textwidth}
\centering
\subfigure[\hspace{0.2cm} log ($\lm = 0$)]{
\label{fig:power-log:l=0-e=005}
\rotatebox{0}{\includegraphics[width = 6.0cm, height = 4.0cm, scale = 1]{power-new_log_l=0_m=100E-2_sm=500E-3_v=395E-1_theta=025_phi=025}}}
\end{minipage}%
\hspace{2.0cm}
\begin{minipage}[t]{0.3 \textwidth}
\centering
\subfigure[\hspace{0.2cm} log ($\lm = 1$)]{
\label{fig:power-log:l=1-e=005}
\rotatebox{0}{\includegraphics[width = 6.0cm, height = 4.0cm, scale = 1]{power-new_log_l=1_m=100E-2_sm=500E-3_v=395E-1_theta=025_phi=025}}}
\end{minipage}
\begin{minipage}[t]{0.3 \textwidth}
\vspace{5.5mm}
\centering
\subfigure[\hspace{0.2cm} exp ($\lm = 0$)]{
\label{fig:power-exp:l=0-e=005}
\rotatebox{0}{\includegraphics[width = 6.0cm, height = 4.0cm, scale = 1]{power-new_exp_l=0_m=100E-2_sm=500E-3_v=395E-1_theta=025_phi=025}}}
\end{minipage}%
\hspace{2.0cm}
\begin{minipage}[t]{0.3 \textwidth}
\vspace{5.5mm}
\centering
\subfigure[\hspace{0.2cm} exp ($\lm = 1$)]{
\label{fig:power-exp:l=1-e=005}
\rotatebox{0}{\includegraphics[width = 6.0cm, height = 4.0cm, scale = 1]{power-new_exp_l=1_m=100E-2_sm=500E-3_v=395E-1_theta=025_phi=025}}}
\end{minipage}%
\caption{\label{fig:power-e=005a} Power spectra for logarithmic and exponential mass functions determined by a spinning particle in
an inspiraling orbit, where $\varepsilon = 0.01$, $\hat{\th} = \hat{\ph} = \pi/4$, $v = 0.395 \, c$, and ${\cal S}/m = 5 \times 10^{-3} \, M_0$.
\vspace{1mm}}
\end{figure*}
Concerning the contribution of MPD spin-gravity coupling to the angular momentum emission rate,
this can be estimated by comparing the $\lm = 0$ and $\lm = 1$ plots in Figure~\ref{fig:ang-t-exp:e=001},
where it follows that the spin interaction reduces the emission rate by about \mbox{5\%}.
In addition, it follows that the exponential mass accretion rate increases $\langle dL^\th/dt \rangle_{\rm rad}$
by roughly \mbox{10\%} over the time interval where the growth rate is in effect.

\subsection{Power spectrum analysis}

\begin{figure*}
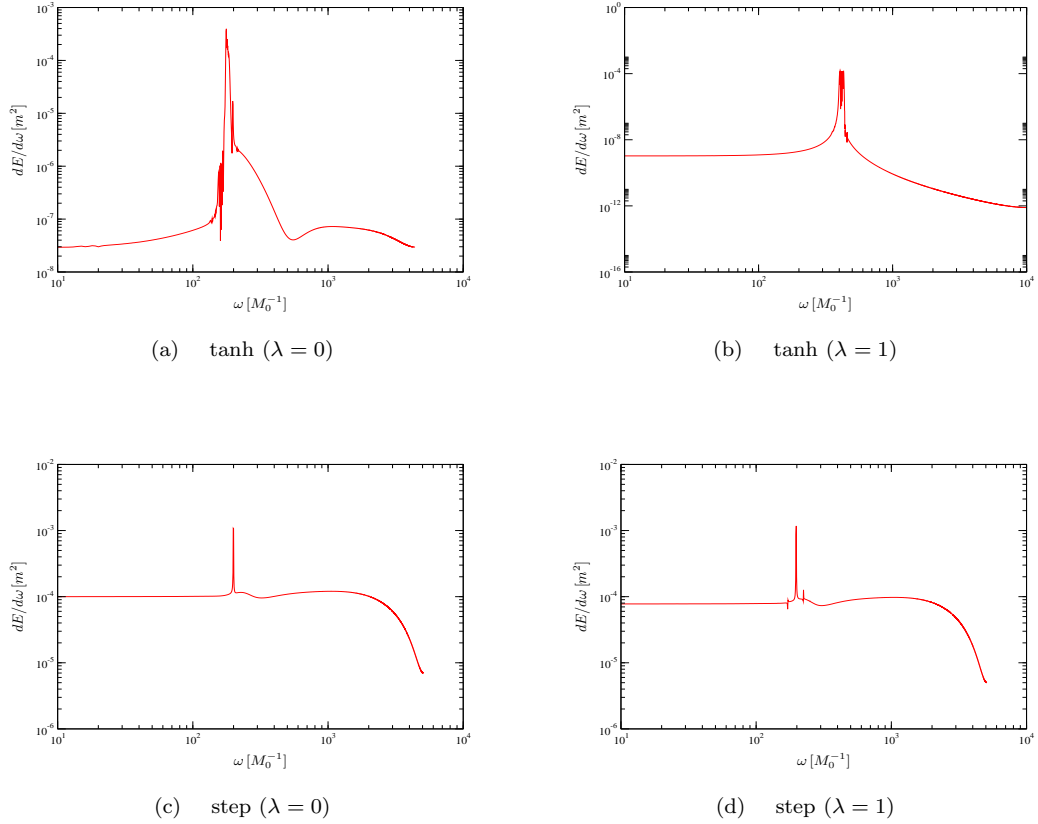

\psfrag{omega [M^(-1)]}[tc][][2.5][0]{$\omega \, [M_0^{-1}]$}
\psfrag{dE/d(omega) [m^2]}[bc][][2.5][0]{$dE/d\omega \, [m^2]$}
\vspace{1mm}
\begin{minipage}[t]{0.3 \textwidth}
\centering
\subfigure[\hspace{0.2cm} tanh ($\lm = 0$)]{
\label{fig:power-tan:l=0-e=005}
\rotatebox{0}{\includegraphics[width = 6.0cm, height = 4.0cm, scale = 1]{power-new_tan_l=0_m=100E-2_sm=500E-3_v=395E-1_theta=025_phi=025}}}
\end{minipage}%
\hspace{2.0cm}
\begin{minipage}[t]{0.3 \textwidth}
\centering
\subfigure[\hspace{0.2cm} tanh ($\lm = 1$)]{
\label{fig:power-tan:l=1-e=005}
\rotatebox{0}{\includegraphics[width = 6.0cm, height = 4.0cm, scale = 1]{power-new_tan_l=1_m=100E-2_sm=500E-3_v=395E-1_theta=025_phi=025}}}
\end{minipage}
\begin{minipage}[t]{0.3 \textwidth}
\vspace{5.5mm}
\centering
\subfigure[\hspace{0.2cm} step ($\lm = 0$)]{
\label{fig:power-stp:l=0-e=005}
\rotatebox{0}{\includegraphics[width = 6.0cm, height = 4.0cm, scale = 1]{power-new_stp_l=0_m=100E-2_sm=500E-3_v=395E-1_theta=025_phi=025}}}
\end{minipage}%
\hspace{2.0cm}
\begin{minipage}[t]{0.3 \textwidth}
\vspace{5.5mm}
\centering
\subfigure[\hspace{0.2cm} step ($\lm = 1$)]{
\label{fig:power-stp:l=1-e=005}
\rotatebox{0}{\includegraphics[width = 6.0cm, height = 4.0cm, scale = 1]{power-new_stp_l=1_m=100E-2_sm=500E-3_v=395E-1_theta=025_phi=025}}}
\end{minipage}%
\caption{\label{fig:power-e=005b} Power spectra for the hyperbolic tangent and step mass functions determined by a spinning particle in
an inspiraling orbit, where $\varepsilon = 0.01$, $\hat{\th} = \hat{\ph} = \pi/4$, $v = 0.395 \, c$, and ${\cal S}/m = 5 \times 10^{-3} \, M_0$.
\vspace{1mm}}
\end{figure*}
An important aspect of this analysis concerns determining the frequency distribution of the power output
associated with the gravitational radiation from the spinning particle.
It is particularly interesting to know the impact of MPD spin gravity coupling on the power spectrum,
and its relative contribution compared to that of purely geodesic motion.
Figures~\ref{fig:power-sch}--\ref{fig:power-tan-stp} describe the power spectra for a spinning test particle
in stable quasi-circular orbit for each of the mass functions considered,
assuming a spin orientation of $\hat{\th} = \hat{\ph} = \pi/4$, where the spectra are defined in units of $m^2$.

For the Schwarzschild mass function with $\lm = 0$, Figure~\ref{fig:power-sch:l=0} shows a single resonance peak at
$\omega \approx 400 \, M_0^{-1}$ with a maximum power of $dE/d\omega \approx 8.8 \times 10^{-3} \, m^2$.
This peak, corresponding to the frequency of the orbit, is common to all power spectrum plots presented here.
By comparison as shown in Figure~\ref{fig:power-sch:l=1}, the contribution of MPD spin gravity coupling has the
effect of generating two well-defined but smaller peaks, with a frequency separation of $\Delta \omega \approx 100 \, M_0^{-1}$.
While there is a slight reduction of maximum power at the principal frequency, there is no fundamental difference
present between the two plots.

The effect of mass accretion imposes distinctive structure on the power spectrum plots.
As shown in Figure~\ref{fig:power-log-exp}, the
the power spectra for the logarithmic and exponential mass functions exhibit similar properties, both for
$\lm = 0$ and $\lm = 1$, particularly in that the principal peak is slightly broader than found in the Schwarzschild case.
This is not surprising given that the mass functions evolve similarly over time.
Comparison of Figures~\ref{fig:power-log:l=0-e=001} and~\ref{fig:power-log:l=1-e=001} for the logarithmic mass function
shows that the MPD spin-gravity coupling also generates the two smaller peaks with the same frequency separation
as found in Figure~\ref{fig:power-sch}, except that the lowest frequency one at $\omega \approx 300 \, M_0^{-1}$
is slightly smaller in height, and that both smaller peaks are slightly distorted.
These properties also present for the exponential mass function case in
Figures~\ref{fig:power-exp:l=0-e=001} and~\ref{fig:power-exp:l=1-e=001}.

The hyperbolic tangent and step mass functions in Figure~\ref{fig:power-tan-stp} generate a power spectra with much greater
complexity than found with the other functions.
This is most evident when the effects of MPD spin-gravity coupling are considered, since at least
four or five smaller peaks can be easily identified in Figures~\ref{fig:power-tan:l=1-e=001}
and~\ref{fig:power-stp:l=1-e=001}.
The only exception in similarity comes from comparing Figures~\ref{fig:power-tan:l=0-e=001} and~\ref{fig:power-stp:l=0-e=001} for
$\lm = 0$, where the former reflects a much greater influence of MPD spin gravity coupling on the corresponding power
spectrum than the latter.

For the case of an infalling spinning particle with $\hat{\th} = \hat{\ph} = \pi/4$, $v = 0.395 \, c$,
and ${\cal S}/m = 5 \times 10^{-3} \, M_0$, the peak structure for the power spectrum changes considerably.
Figures~\ref{fig:power-e=005a} and~\ref{fig:power-e=005b} show that only one resonance
largely appears and is shifted towards the viscinity of $\omega \approx 200 \, M_0^{-1}$ for all plots.
The only exception to this pattern is Figure~\ref{fig:power-tan:l=1-e=005}, whose resonance peak remains at around
$\omega \approx 400 \, M_0^{-1}$.
For all cases, however, the power output at resonance drops by about an order of magnitude
compared to the corresponding sets for stable orbits.


\section{Observational Possibilities}

The essential motivation for examining the dynamical behaviour of spinning particles in Vaidya space-time
is to determine the predicted gravitational waveforms that may, in principle, become observable by detectors
such as LIGO or LISA.
However, the likelihood of actually observing the effects of radiation-induced mass accretion in this manner,
both on conceptual and practical grounds, seems negligible when compared to the behaviour determined by
a Schwarzschild black hole.
The main reason for this is because the amount of mass accretion required over a reasonable observation time
is many orders of magnitude larger than allowed for by the Eddington luminosity limit for physically realistic mass accretion models.
Further to this point is that because the central mass is non-rotating, the system lacks the presence
of spin angular momentum to drive up the mass accretion rate to a more robust level.

To better illustrate the difficulty of observing spherically symmetric mass accretion in the waveforms,
the Eddington limit imposes a mass accretion rate of \cite{Heyl}
\be
\dot{M} & = & 3 \times 10^{-22} \, \gamma^{-1} \, \lt({M_0 \over M_\odot}\rt),
\label{Mdot}
\ee
in geometric units, where $\gamma \approx 0.1$ is the energy release efficiency of the
outgoing photon flux to counterbalance the radiation mass accretion.
To correct for the condition that only a small fraction of the infalling photons get radiated away,
the mass accretion rate is then $dM/dt = \lt(1 - \gamma\rt) \dot{M}$.
Therefore, the time required to observe a central mass increase of $\Delta M = \varepsilon \, M_0$ is
\be
\Delta t & = & {\Delta M \over dM/dt} \ = \ 3.7 \times 10^{20} \, \varepsilon \, M_\odot
\nn
& = & 5.8 \times 10^7 \, \varepsilon \, {\rm \ years}.
\label{delta-t}
\ee
To then detect within the gravitational waveforms a 1\% increase in the central mass requires an observation time
on the order of $10^5$ years.
Similarly, a one-year observation time can yield, at best, an observable mass increase of only
$\Delta M \approx 10^{-8} \, M_0$, which must be negligibly small for the design capabilities of LIGO or LISA.
It is uncertain at present whether such a small effect may become reasonably detectable with the advent of
future gravitational wave observatories.
Nevertheless, the usefulness of this investigation comes from providing a theoretical testing ground for these
same mass accretion models by comparing the predicted waveforms with actual signals.

Given the power spectrum plots of Figures~\ref{fig:power-sch}--\ref{fig:power-e=005b}, it is important to
know the size of central mass that is required for the resonance to fall within the frequency detection bands of
LIGO and LISA.
This can be determined by letting $M_0 = \alpha \, M_\odot$ for some unspecified parameter $\alpha$ and evaluating
the resonance peak at around $\omega \approx 400 \, M_0^{-1}$ in terms of $M_\odot^{-1} = 2.03 \times 10^5$ Hz.
It then follows that
\be
\omega & \approx & 8.12 \times 10^7 \, \alpha^{-1} \, {\rm \ Hz}.
\label{omega-Hz}
\ee
Assuming that the spinning point particle is a solar mass Kerr black hole or rapidly rotating neutron star,
$\alpha = 10^3$ and $\omega \approx 8.12 \times 10^4$ Hz, which is about two orders of magnitude beyond LIGO's
upper limit in its frequency band, but may perhaps be open for observation by a next-generation interferometer.
To fall well within LIGO's bandwidth then requires that $\alpha = 10^5-10^6$, which implies that the orbiting
point particle must be an intermediate-sized Kerr black hole with $m = 10^3-10^4 \, M_\odot$.
In order to satisfy LISA's bandwidth, it is necessary that $\alpha = 10^8-10^9$, with $m = 10^6-10^7 \, M_\odot$.
On the surface, it seems implausible to expect to find in abundance binary systems composed of {\em two} supermassive black holes.
Nonetheless, it may not be completely outside the realm of possibility, since there appears to be no obvious reason
why supermassive black hole binaries should not be found within the known Universe.


\section{Conclusion}

This paper investigated the dynamics of a classical spinning point particle in orbit around
a gravitationally collapsing central mass due to radiation accretion, as described by the Vaidya metric.
It is demonstrated that the Mathisson-Papapetrou-Dixon equations have a discernable influence on the
motion of the spinning particle.
A comprehensive numerical analysis of the particle's orbital kinematics and dynamics is performed
for various mass functions which satisfy the weak energy condition.
In addition, a leading-order description of the gravitational waveforms generated by the spinning particle is
presented, along with a power spectrum analysis.
For an observer positioned along the orbital plane, it is shown that the $\times$ polarization mode
captures most clearly the properties of MPD spin-gravity coupling, with an amplitude of around 5\% compared to
the $h_+$ polarization mode for the choice of initial conditions considered.
However, it is also shown that any effects for a realistic mass function with an accretion rate satisfying the Eddington
luminosity limit must be very small.

There are a number of potential considerations for future work based on this paper.
Because the waveforms are determined using the leading-order quadrupole moment formula only,
it may be necessary to calculate more precisely defined waveforms based on black hole perturbation theory.
It should be noted, though, that such an endeavour may yield only a minimum level of correction, given that
a realistic mass accretion rate is very small in the first place.
Likely a more meaningful advance of this work is to allow for rotation of the central mass, as described by the Kerr-Vaidya
metric \cite{Kerr-Vaidya}, and repeat the analysis presented here.
Another interesting possible future development is to consider the spinning particle dynamics around
a star undergoing a supernova explosion.
While there is some complexity involved given that a supernova evolves in a highly asymmetric fashion, the
fact that a sudden change of space-time curvature that follows from the event should have a significant impact
on the orbital properties of the spinning particle, with consequences for the gravitational radiation emitted.
These and other possible developments may be considered in the future.


\section{Acknowledgments}

The author is immensely grateful to Bahram Mashhoon for the many suggestions
offered to enhance the quality of this paper, and is thankful to him and the
Department of Physics and Astronomy at the University of Missouri-Columbia
for hospitality shown during this time.
He gratefully acknowledges Scott Hughes of MIT for helpful discussions and insights offered.
The author also thanks Charles Wang of Lancaster University (now at the University of Aberdeen)
for introducing him to the study of Dixon particles, and for his overall encouragement of this project.

\end{document}